\documentclass[12pt]{iopart}
\usepackage{amsfonts}
\usepackage{amssymb}
\usepackage{graphicx,xcolor}%
\usepackage[hang,raggedright]{subfigure}

\def\N#1{\left|\!\left|{#1}\right|\!\right|} 
\def\>{\rangle}\def\<{\langle} \def\sH{\mathcal{H}}
\newcommand{\bea}{\begin{eqnarray}}
  \newcommand{\F}{\mathsf{F}} 

  \newcommand{\eea}{\end{eqnarray}}
\def\ent{\mathrm{ent}}
\def\sS{\mathcal{M}}
\def\id{{\rm id}}

\def\N#1{\left|\!\left|{#1}\right|\!\right|} 
\def\>{\rangle}\def\<{\langle} \def\sH{\mathcal{H}}

\newcommand{\ket}[1]{| #1 \rangle}

\newtheorem{theorem}{Theorem}

\newtheorem{definition}[theorem]{Definition}

\newtheorem{lemma}[theorem]{Lemma}

\newenvironment{proof}[1][Proof]{\noindent\textbf{#1} }{\ \rule{0.5em}{0.5em}}

        \def\cB{{\cal B}}

\def\cD{{\cal D}}        \def\cE{{\cal E}}

        \def\cH{{\cal H}}

\def\cM{{\cal M}}        \def\cN{{\cal N}}

\def\cV{{\cal V}}

\def\0{{\mathbf{0}}}
\def\1{{\mathbf{1}}}
\def\2{{\mathbf{2}}}
\def\3{{\mathbf{3}}}
\def\4{{\mathbf{4}}}
\def\5{{\mathbf{5}}}
\def\6{{\mathbf{6}}}
\def\7{{\mathbf{7}}}
\def\8{{\mathbf{8}}}
\def\9{{\mathbf{9}}}

\def\bbI{{\mathbb{I}}}

\def\be{\begin{equation}}
\def\ee{\end{equation}}
\def\bea{\begin{eqnarray}}
\def\eea{\end{eqnarray}}
\def\reff#1{(\ref{#1})}
\def\eps{\varepsilon}

\begin{document}

\title{The apex of the family tree of protocols: Optimal rates and resource inequalities}
\author{Nilanjana Datta and Min-Hsiu Hsieh}
\address{Statistical Laboratory, University of Cambridge, Wilberforce Road, Cambridge CB3 0WB, UK}
\ead{n.datta@statslab.cam.ac.uk}
\ead{minhsiuh@gmail.com}
\begin{abstract}
We establish bounds on the {maximum} entanglement gain and {minimum} quantum communication cost of the Fully Quantum Slepian-Wolf protocol in the one-shot regime, which is considered to be at the apex of the existing family tree in Quantum Information Theory. These quantities, which are expressed in terms of smooth min- and max-entropies, reduce to the known rates of quantum communication cost and entanglement gain in the asymptotic i.i.d.~scenario. We also provide an explicit proof of the optimality of these asymptotic rates. We introduce a resource inequality for the one-shot FQSW protocol, which in conjunction with our results, yields achievable one-shot rates of its children protocols. In particular, it yields bounds on the one-shot quantum capacity of a noisy channel in terms of a single entropic 
quantity, unlike previously bounds. We also obtain an explicit expression for the achievable rate for one-shot state redistribution.
\end{abstract}
\maketitle

%%%%%%%%%%%%%%%%%%%%%
% Introduction
%%%%%%%%%%%%%%%%%%%%%
\section{Introduction}
An important problem in Quantum Information Theory is the evaluation of optimal rates of various information processing tasks. These include data compression \cite{Schumacher95,JS94}, information transmission \cite{Holevo73,Holevo98,SW97,Devetak03,Lloyd96,Shor02}, entanglement manipulation \cite{BBPS96,LP99PRL,DW04PRL,DW05}, state merging \cite{HOW05Nature,HOW06CMP}, channel simulation \cite{BSST02IT,BDHSW09} and a host of other protocols \cite{BSST02IT,DHW04PRL,DHW08IT,HHHLT01,Holevo02,HDW08IT,HW10IT,HW10IT1,HW10IT2}. Initially, all these different protocols were considered to be unrelated and each one of them was studied individually. That was until the authors of \cite{DHW04PRL} proposed two new protocols, namely, the ``mother'' and the ``father'' protocols, from which five of the previously studied protocols could be generated by direct application of teleportation and superdense coding. More precisely, entanglement distillation \cite{DW04PRL,DW05}, noisy teleportation \cite{DHW04PRL} and noisy superdense coding \cite{HHHLT01} could be generated from the ``mother'', whereas the transmission of quantum information through a quantum channel \cite{Devetak03,Lloyd96,Shor02}, and the entanglement-assisted transmission of classical information through a quantum channel \cite{BSST02IT,Holevo02,HDW08IT} could be generated from the ``father''.

Within the {\em{resource framework}}, developed in~\cite{DHW04PRL,DHW08IT}, Quantum Shannon Theory can be ultimately viewed as the study of inter-conversions between non-local information-processing resources. These resources can be classified either as static (e.g., shared randomness or entanglement) or dynamic (e.g., communication channels).  Further, they may be either noisy or noiseless, and finite or asymptotic. Noiseless resources are the fundamental ingredients of information theory because they can be employed to achieve essential tasks (e.g., information transmission, teleportation etc.) {\em{perfectly}}, that is without any error. The basic unit of a noiseless quantum static resource is represented by an EinsteinÐPodolskyÐRosen (EPR) pair (an ebit), whereas the corresponding dynamic one is a noiseless single-qubit channel. At the most fundamental level, optimal rates of protocols hence characterize the amount of noiseless resources which can be extracted from a given noisy one. The ``mother'', which is a quantum communication-assisted entanglement distillation protocol, is a parent of protocols which involve ``static'' resources, whereas the ``father'', which is an entanglement-assisted quantum communication protocol, is a parent of  protocols which involve``dynamic'' resources.

In 2006, yet another protocol, called the Fully Quantum Slepian-Wolf (FQSW) was proposed \cite{ADW09PRSA}, which achieved the remarkable feat of unifying the family tree mentioned above. It is hence also referred to as ``{\em{the mother of all protocols}}''. It is a generalization of the ``mother'' protocol since, in addition to quantum communication-assisted entanglement distillation, it accomplishes state transfer from the sender to the receiver. Moreover the FQSW can also be transformed into the ``father'' protocol by employing Schmidt symmetry \cite{ADW09PRSA, patricknotes}. In addition, the FQSW protocol, can be used as a primitive for the following important protocols: state merging \cite{HOW05Nature,HOW06CMP}, simulation of channels (a fully quantum reverse Shannon theorem) \cite{BSST02IT,BDHSW09,BCR09}, quantum communication through broadcast channels \cite{YHD06,DHL10IT}, distributed compression \cite{ADW09PRSA} and state redistribution \cite{YD09IT,DY08PRL,YBW08PRA}. It is hence evident that the FQSW protocol is at the heart of Quantum Shannon Theory. The FQSW protocol derives its name from its applicability to distributed compression, a problem which was solved in the classical case by Slepian and Wolf \cite{SW71}. It is also referred to as state transfer \cite{patricknotes} or the merging mother \cite{Oppenheim08}. Note, however, that the family tree is not exhaustive since it does not cover all information-theoretic protocols. Also the tree structure is not unique, since, for example, it is possible to obtain the FQSW from state merging via superdense coding \cite{Oppenheim08}.

Optimal rates of quantum information-processing tasks were originally evaluated in the so-called {\em{asymptotic~i.i.d.~scenario}}, i.e., in the limit of asymptotically many uses of the underlying resources, under the assumption that there was no correlation between successive uses. In other words, quantum channels employed in the protocols were assumed to be memoryless and entanglement resources were assumed to consist of states which were multiple copies (and hence tensor products) of a given entangled state.

In real-world applications, however, this assumption and the consideration of the asymptotic scenario is not necessarily justified. A more general theory of quantum information-processing protocols is obtained instead in the so-called {\em{one-shot scenario}} \cite{BCR09,WR10,RR10DC, RR10,BD10IT,Dupuis2010,BD10,BD10JMP,BertaThesis,DupuisThesis} in which resources are considered to be finite and the information-processing tasks are required to be achieved only up to a finite accuracy. This also corresponds to the scenario in which experiments are performed since channels and entanglement resources available for practical uses are typically finite and correlated, and transformations can only be achieved approximately. The fact that the one-shot scenario is more general than the asymptotic i.i.d.~one is further evident from the fact that optimal rates of protocols in the latter can directly be obtained from the corresponding one-shot rates. Further, one-shot rates also yield asymptotic rates for protocols involving correlated resources via the Quantum Information Spectrum method (see e.g.~\cite{q-spectra, bowen} and references therein).
%See e.g. \cite{BD10IT, qdistil, wang?}).

In this paper we focus on the one-shot FQSW protocol and evaluate upper and lower bounds on its optimal rates. In this protocol, one starts with a {\em{single copy}} of a tripartite pure state $\ket{\psi}^{ABR}$, where the system $A$ is with Alice, $B$ is with Bob and $R$ denotes the purifying reference system. The aim is for Alice to transfer the $A$-part of the state to Bob and at the same time generate entanglement with him. Alice and Bob can both do local operations on systems in their possession and Alice can send qubits to Bob. The minimum quantum communication cost, i.e., the minimum number of qubits that Alice needs to send to Bob in order to achieve the state transfer (up to a given finite accuracy) and the maximum resulting entanglement gain are referred to as the optimal rates of the one-shot FQSW protocol. Since this protocol is the one-shot version of the ``mother of all protocols'', it can be viewed as the most basic building block of Quantum Shannon Theory and is at the {\em{apex of the family tree of protocols}}.

{{The one-shot FQSW was first introduced in \cite{ADW09PRSA} and studied in \cite{BCR09, {DupuisThesis}}. Moreover a classical-quantum version of it was treated in \cite{RR10DC}}}. In \cite{BCR09} an expression for the achievable rates was obtained in terms of an unsmoothed min-entropy. However, the optimality of these rates or the analysis of the asymptotic case, was not addressed. In contrast, we obtain both lower and upper bounds on the optimal rate of one-shot FQSW. Our achievable rates are expressible in terms of smooth min- and max-entropies \cite{RennerThesis, TCR09IT, KRS09IEEEIT, TCR10IEEEIT} which have the advantage of directly yielding the known {{achievable rates}} of the FQSW in the asymptotic i.i.d.~scenario (which are expressed in terms of the mutual information) \cite{ADW09PRSA}. Moreover, our one-shot results can be used to prove that these asymptotic rates of quantum communication cost and entanglement gain are indeed optimal \cite{HOW06CMP}. Further we introduce a resource inequality for the one-shot FQSW protocol. This leads to resource inequalities and achievable rates for the one-shot ``mother'' and ``father'' protocols and their children. In this paper we list some of these, a more exhaustive study of all the children protocols being deferred to a forthcoming paper.  In particular, we obtain upper and lower 
bounds on the one-shot quantum capacity of a noisy channel in terms of the {\em{same}} entropic quantity, namely a smooth max-entropy,
unlike previously obtained bounds \cite{BD10IT}. As shown in \cite{jonathan}, the FQSW protocol can be used as a primitive for state redistribution. Consequently, our results on the one-shot FQSW can be used to obtain an expression for the achievable rate for one-shot state redistribution. A more detailed analysis of this will be presented in \cite{jonathan}.
The smooth min- and max-entropies appearing in our theorems have interesting properties and satisfy a series of useful inequalities
(see e.g. Appendix A). These relations and a one-shot decoupling lemma are the main ingredients of our proofs.

The paper is organized as follows. We start with some definitions and notations in Section \ref{SEC_DEF}. In Section \ref{SEC_MAIN}, we state our main results on the one-shot FQSW, the optimality of the rates of quantum communication cost and entanglement gain in the asymptotic i.i.d. scenario, and the bounds on the one-shot quantum capacity of a noisy channel. In Section \ref{child} we introduce resource inequalities for the one-shot FQSW and some of its children protocols. Theorem~\ref{theorem5}, Theorem~\ref{thmconverse}, Theorem~\ref{asymp} and Theorem~\ref{one-qcap}, and the one-shot resource inequalities of Section \ref{child}, constitute the main results of this paper. Proofs of Theorem~\ref{theorem5} and Theorem~\ref{thmconverse} are given in Section \ref{SEC_PROOF}, while the one-shot decoupling theorem, which is employed in the proof of Theorem~\ref{theorem5}, is proved in Appendix B. In Appendix A we list various entropic inequalities which we use.

%%%%%%%%%%%%%%%%%%%%%%%%%%%%%%%%%%
% Notations and Definitions
%%%%%%%%%%%%%%%%%%%%%%%%%%%%%%%%%%
\section{Notations and Definitions}
\label{SEC_DEF}%
Let $\cB(\cH)$ denote the algebra of linear operators acting on a finite-dimensional Hilbert space $\cH$, and let ${\cD}({\cH})\subset\cB(\cH)$ denote the set of positive operators of unit trace (states) on $\cH$. Furthermore, let ${\cD}_{\leq}({\cH})$ denote the set of subnormalized states. Throughout this paper, we restrict our considerations to finite-dimensional Hilbert spaces and denote the dimension of a Hilbert space ${\cH}_A$ by $|A|$.

For any given pure state $|\psi\> \in {\cH}$, we denote the projector $|\psi\>\<\psi|$ simply as $\psi$. For a positive semi-definite operator $\omega^{AB} \in \cB(\cH_A \otimes \cH_B)$, let $\omega^{A} := \tr_B \omega^{AB}$ denote its restriction to the subsystem $A$. For given orthonormal bases $\{|i^{A}\rangle\}_{i=1}^d$ and
$\{|i^{B}\rangle\}_{i=1}^d$ in isomorphic Hilbert spaces
${\cal{H}}_{A}\simeq{\cal{H}}_B\simeq{\cH}$ of dimension $d$, we define a
maximally entangled state (MES) of Schmidt rank $d$ to be
\begin{equation}\label{MES-m}
|\Phi\>^{AB}= \frac{1}{\sqrt{d}} \sum_{i=1}^d |i^A\rangle\otimes |i^B\rangle.
\end{equation}
Let $\bbI_{A}$ denote the identity operator in ${\cal{B}}({\cal{H}}_A)$, and let $\tau^A := \bbI_A/|A|$ denote the completely mixed state in
 ${\cal{D}}({\cal{H}}_A)$.

In the following we denote a completely positive trace-preserving (CPTP)
map $\cE: \cB({\cH}_A) \mapsto \cB({\cH}_B)$ simply as
$\cE^{A\to B}$. Similarly, we denote an isometry $U: {\cH}_A \mapsto {\cH}_B \otimes {\cH}_C$ simply as $U^{A\to BC}$.

The trace distance between two operators $A$ and $B$ is given by
\begin{equation}\nonumber
\N{A-B}_1 := \tr\bigl[\{A \ge B\}(A-B)\bigr] - \tr\bigl[\{A <  B\}(A-B)\bigr],
\end{equation}
where $\{A\ge B\}$ denotes the projector on the subspace where the operator $(A-B)$ is non-negative, and
$\{A<B\}:=\bbI-\{A\ge B\}$. The fidelity of two states $\rho$ and $\sigma$ is defined as
\begin{equation}\label{fidelity-aaa}
F(\rho, \sigma):= \tr \sqrt{\sqrt{\rho} \sigma \sqrt{\rho}}
=\N{\sqrt{\rho}\sqrt{\sigma}}_1.
\end{equation}
Note that the definition of fidelity can be naturally extended to subnormalized states.
The trace distance between two states $\rho$ and $\sigma$ is related
to the fidelity $F(\rho, \sigma)$ as follows (see e.~g.~\cite{NC00}):
\begin{equation}
  1-F(\rho,\sigma) \leq \frac{1}{2} \N{\rho -
    \sigma}_1 \leq \sqrt{1-F^2(\rho, \sigma)},
\label{fidelity}
\end{equation}
where we use the notation $F^2(\rho, \sigma) = \bigl(F(\rho,\sigma)
\bigr)^2$. For $\rho \in {\cD}({\cH})$ and $\sigma \in {\cD}_{\leq}({\cH})$ we
also use the quantity
\be\label{crho}
C(\rho, \sigma) := \sqrt{ 1 - F^2(\rho, \sigma)},
\ee
which was introduced in \cite{Gilchrist} and proved to be a metric. It is monotonic under any CPTP map $\cE$, i.e.,
\be
C(\rho, \sigma) \ge C(\cE(\rho), \cE(\sigma)).
\label{mono}
\ee
Moreover, if $\rho,\sigma\in\cD(\cH)$, then
\be\label{CC}
C(\rho, \sigma) \le \sqrt{\N{\rho-\sigma}_1}.
\ee
This follows by noting that $C(\rho, \sigma)$ is a special case of the purified distance $P(\rho, \sigma)$ (introduced in \cite{TCR10IEEEIT}), which satisfies these properties. We use the following lemma,
\begin{lemma}[Gentle Measurement Lamma \cite{gentle1,gentle2}]\label{gentle}
For a state $\rho\in\cD(\cH)$ and operator $0\leq\Lambda\leq \bbI$, if $\tr\Lambda\rho\geq 1-\delta$, then
\[
\|\rho-\sqrt{\Lambda}\rho\sqrt{\Lambda}\|_1\leq2\sqrt{\delta}.
\]
The same holds if $\rho$ is a subnormalized density operator.
\end{lemma}

The results in this paper involve various entropic quantities. The von Neumann entropy of a state $\rho^A\in\cD(\cH_A)$ is given by $H(A)_{\rho} = - \tr \rho^A \log \rho^A$. Throughout this paper we take the logarithm to base $2$. 
%For positive semi-definite operators $\rho$ and $\sigma$, the quantum relative R\'enyi entropy of order $\alpha$, with $0\leq\alpha<1$, is defined as (see e.g. \cite{OP, hayashibook}):
%\be
%S_\alpha(\rho||\sigma):= \frac{1}{\alpha-1}\log \tr \rho^{\alpha}\sigma^{1-\alpha}.
%\label{renyi}
%\ee
For any state $\rho^{AB} \in \cD(\cH_{A}\otimes\cH_B)$, the coherent information $I(A\rangle B)_\rho$ and the quantum mutual information $I(A:B)_\rho$ are defined respectively as
\begin{eqnarray}
I(A\rangle B)_\rho&:=& H(B)_\rho-H(AB)_\rho \label{coherent}\\
I(A:B)_\rho&:=& H(A)_{\rho} +  H(B)_{\rho}- H(AB)_{\rho}. \label{mutual}
\end{eqnarray}

In addition to the above entropic quantities, we make use of the following generalized relative entropy quantity, referred to as the max-relative entropy, introduced in \cite{Datta09IT}:
\begin{definition}\label{DEF_MAX_RELATIVE}
The max-relative entropy of two operators $\rho\in\cD_{\leq}(\cH)$ and $\sigma\in\cB(\cH)$, $\sigma\geq 0$, is defined as
\begin{equation}
D_{\max}(\rho||\sigma):= \log\min\{\lambda:\rho\leq\lambda\sigma\}.
\end{equation}
\end{definition}
We also use the following conditional min- and max-entropies defined in \cite{RennerThesis, TCR09IT, TCR10IEEEIT}:
\begin{definition}
Let $\rho^{AB}\in\cD_{\leq}(\cH_{A}\otimes\cH_B)$. The min-entropy of $A$ conditioned on $B$ for the state $\rho^{AB}$ is defined as
\[
H_{\rm min}(A|B)_\rho=\max_{\sigma^B\in\cD(\cH_B)}\left[-D_{\max}(\rho^{AB}||\bbI_A\otimes\sigma^B)\right].
\]
\end{definition}

\begin{definition} For any $0\leq \eps\leq 1$ and $\rho\in\cD(\cH)$, we define the $\eps$-ball around $\rho$ as follows
\[
\cB^\eps(\rho)=\{{\overline{\rho}}\in\cD_{\leq}(\cH):F^2({\overline{\rho}}, \rho)\geq1-\eps^2\}.
\]
Note that if ${\overline{\rho}}\in \cB^\eps(\rho)$ then $C({\overline{\rho}}, \rho) \le \eps$.
\end{definition}

\begin{definition}
Let $0\leq \eps\leq 1$ and $\rho^{AB}\in\cD(\cH_{A}\otimes\cH_B)$. The $\eps$-smooth min-entropy of $A$ conditioned on $B$ for the state $\rho^{AB}$ is defined as
\[
H_{\rm min}^\eps(A|B)_\rho =\max_{\overline{\rho}^{AB}\in\cB^\eps(\rho^{AB})} H_{\min}(A|B)_{\overline{\rho}}.
\]
\end{definition}
We also use the max-entropy which is defined in terms of the min-entropy via the following duality relation \cite{TCR09IT, TCR10IEEEIT, KRS09IEEEIT}:
\begin{definition}\cite{TCR10IEEEIT}\label{LEM_DUALITY}
Let $\rho^{AB}\in\cD(\cH_{A}\otimes\cH_B)$ and let $\rho^{ABC}\in\cD(\cH_{A}\otimes\cH_B\otimes\cH_C)$ be an arbitrary purification of $\rho^{AB}$. Then for any $0\leq \eps\leq 1$,
\be\label{EQ_dual}
H_{\rm max}^\eps(A|C)_\rho := - H_{\rm min}^\eps(A|B)_\rho.
\ee
\end{definition}
When $\rho^{AB}$ is a pure state, then
\be
H_{\rm min}^\eps(A|B)_\rho = - H_{\rm max}^\eps(A)_\rho.
\ee
For any state $\rho^{AB} \in \cD(\cH_{A}\otimes\cH_B)$ the $\eps$-smooth conditional max-entropy can be equivalently expressed as \cite{TCR10IEEEIT, KRS09IEEEIT}
\be
H_{\rm max}^\eps(A|B)_\rho := \min_{{\overline{\rho}}^{AB}\in
\cB^\eps(\rho^{AB})}
H_{\rm max}(A|B)_{\overline{\rho}}\label{EQ_smooth_hmax},
\ee
where
\be\label{star}
H_{\rm max}(A|B)_{\overline{\rho}}= \max_{\sigma^B \in \cD(\cH_B)}2 \log
F\bigl({\overline{\rho}^{AB}},\bbI_A \otimes\sigma^B\bigr).
\ee
Moreover, for any $\rho^A \in \cD_{\leq}(\cH_{A})$,
\be
H_{\rm max}(A)_\rho = 2 \log \tr {\sqrt{\rho^A}}.\label{EQ_S_half}
\ee

Other than the conditional min- and max-entropies, we also require the entropic quantity defined below.
\begin{definition}
 Given any $\rho^{AB}\in\cD_{\leq}(\cH_A\otimes\cH_B)$, we define
\begin{equation}\label{ent_H0}
H_0(A|B)_\rho=\max_{\sigma^B\in\cD(\cH_B)}\log\tr\Pi_{\rho^{AB}}(\bbI_A\otimes\sigma^B)
\end{equation}
where  $\Pi_{\rho^{AB}}$ denotes the projector onto the support of $\rho^{AB}$. For any $0\leq\eps\leq 1$, define
\begin{equation}\label{ent_Ho_smooth}
\widetilde{H}^\eps_0(A|B)_\rho:=\min_{\overline{\rho}^{AB}\in\cB^\eps(\rho^{AB})}H_0(A|B)_{\overline{\rho}}.
\end{equation}
\end{definition}
When the system $B$ is trivial, we have
\[
H_0(A)_\rho =\log\tr\Pi_{\rho^A},
\]
where $\Pi_{\rho^A}$ denotes the projector onto the support of $\rho^A$.

We also use the following entropic quantity which is obtained by a 
different smoothing of $H_0(A)_{\rho}$. For any $0\leq \eps\leq1$ and any $\rho^{A}\in\cD(\cH_A)$, define
\begin{equation}
H_0^{\eps}(A)_\rho:=\min_{0\leq Q \leq \bbI_A\atop{\tr Q\rho^A\geq 1-\eps}}H_0(A)_{\rho_Q^A}
\end{equation}
where $\rho_Q^A$ is defined as
\begin{equation}\label{EQ_rho_Q}
\rho_Q^A:=\sqrt{Q}\rho^A\sqrt{Q}.
\end{equation}
It follows from the Gentle Measurement lemma (Lemma~\ref{gentle}) that $\|\rho_Q^A-\rho^A\|_1\leq 2\sqrt{\eps}$ and simple calculation gives $\rho_Q^A\in\cB^{\delta}(\rho^A)$, where $\delta=\sqrt{4\sqrt{\eps}-4\eps}$. Various properties of the entropies defined above, which we employ in our proofs, are given in \ref{APPENDIX_A}.

%%%%%%%%%%%%%%%%%%%%%%%
% Main Results
%%%%%%%%%%%%%%%%%%%%%%%
\section{Main Results}%
\label{SEC_MAIN}%
\subsection{Optimal rates for the One-Shot FQSW Protocol}
In the one-shot FQSW protocol, one starts with a {\em{single copy}} of a tripartite pure state $\ket{\psi}^{ABR}$, where the system $A$ is with Alice, $B$ is with Bob, and $R$ denotes the reference system. The aim is for Alice to transfer the $A$-part of the state to Bob and at the same time generate entanglement with him. The protocol is referred to as an $\eps$-error one-shot FQSW protocol, if for any given $0<\eps \leq 1$, the error in achieving this aim is at most $\eps$.

Any $\eps$-error FQSW protocol for $\ket{\psi}^{ABR}$ can be assumed to have the following form: Alice does local operations on the system $A$ and sends a quantum system (i.e., qubits) to Bob, who then performs local operations on the quantum systems in his possession. The final state of the protocol has to be $\eps$-close (in a sense specified below) to the state $\Phi^{A_1B_1}\otimes \psi^{B'BR}$, where the subscripts have been chosen to denote which systems are in whose possession (i.e., $B',B$ and $B_1$ are with Bob and $A_1$ is with Alice), $\Phi^{A_1B_1}$ is a maximally entangled state of size $|A_1|$, and $\psi^{B'BR}$ is identical to the initial tripartite state $\psi^{ABR}$ but with the system $A$ now in Bob's possession. Consequently, the initial entanglement between $A$ and $R$ has been transferred to Bob.

The following two theorems, which together give upper and lower bounds on the minimum quantum communication cost and the maximum entanglement gain of an $\eps$-error one-shot FQSW protocol, constitute the main results of this section.
\begin{theorem}[{\bf{Achievability}}]
\label{theorem5}
Fix $0<\eps\leq 1$. Then for any tripartite pure state $\psi^{ABR}$, there exists an $\eps$-error one-shot FQSW protocol with an entanglement gain $e^{(1)}_{\eps}$ and a quantum communication cost $q^{(1)}_{\eps}$ bounded respectively by
\begin{eqnarray}
e^{(1)}_{\eps}&\geq & \frac{1}{2}\left[H^{\delta}_0(A)_\psi+H_{\rm min}^{\delta}(A|R)_{\psi}\right]+\log\delta'\label{e-achieve}\\
q^{(1)}_{\eps}&\leq & \frac{1}{2}[H_{\rm 0}^{\delta}(A)_\psi- H_{\rm min}^{\delta}(A|R)_\psi]-\log\delta'\label{q-achieve}
\end{eqnarray}
for some $\delta>0$ such that $\eps=2\sqrt{5\delta'}+2\sqrt{\delta}$ and $\delta'=\delta+\sqrt{4\sqrt{\delta}-4\delta}$.
\end{theorem}
The proof of this theorem relies on the one-shot decoupling theorem (Theorem \ref{thm_decoupling}) which is proved in Appendix B. 

\begin{theorem}[{\bf{Converse}}]
\label{thmconverse}
Fix $0<\eps\leq1$. Then given a tripartite pure state $\psi^{ABR}$, the quantum communication cost $q^{(1)}_{\eps}$ and the entanglement gain $e^{(1)}_{\eps}$ of any $\eps$-error one-shot FQSW protocol satisfies the following bounds:
\begin{eqnarray}
q^{(1)}_{\eps}&\geq& \frac{1}{2}
\left[H_{\rm min}^{\eps}(A)_\psi-
H_{\rm min}^{{\overline{\eps}}}(A|R)_\psi\right] -\log\frac{\sqrt{2}}{\eps} \label{lbdq}\\
e^{(1)}_\eps &\leq& q^{(1)}_\eps + H_{\rm min}^{{\overline{\eps}}}(A|R)_\psi +\log\frac{2}{\eps^2} \label{upe1}
\end{eqnarray}
where ${{\overline{\eps}}:}= 3( \eps + \sqrt{3\sqrt{\eps}})$.
\end{theorem}
The proofs of Theorem~\ref{theorem5} and Theorem~\ref{thmconverse} are given in Section~\ref{SEC_PROOF}.

%%%%%%%%%%%%%%%%%%%%%%%%%%%%%%
% FQSW in the asymptotic i.i.d. scenario
%%%%%%%%%%%%%%%%%%%%%%%%%%%%%%
\subsection{Optimality of the rates in the asymptotic~i.i.d.~scenario}
Here we prove how the known achievable rates for the FQSW in the asymptotic i.i.d. scenario can be recovered from theorems stated above. We also prove explicitly that these rates are indeed optimal \cite{HOW06CMP}.

Let $\psi_n^{ABR}:=(\psi^{ABR})^{\otimes n}$ and let ${{q}^{(1)}_{\eps, n}}$ and ${e}^{(1)}_{\eps, n}$ respectively denote the quantum communication cost and entanglement gain of an $\eps$-error one-shot FQSW for the state $\psi_n^{ABR}$. Then the optimal rates of quantum communication and entanglement gain in the asymptotic~i.i.d.~scenario can be respectively defined in terms of ${q}^{(1)}_{\eps, n}$ and ${e}^{(1)}_{\eps, n}$ as follows:
\be\label{infq}
q^\infty:= \lim_{\eps \to 0}\lim_{n\to \infty}\frac{1}{n}{{q}^{(1)}_{\eps, n}}
\ee
and
\be\label{infe}
e^\infty:= \lim_{\eps \to 0}\lim_{n\to \infty}\frac{1}{n}{{e}^{(1)}_{\eps, n}}.
\ee
\begin{theorem}\label{asymp}
The optimal rates of quantum communication cost and entanglement gain for the FQSW protocol for a tripartite pure state $\psi^{ABR}$ in the asymptotic~i.i.d.~scenario are respectively given by the following: \be\label{arates}
q^\infty = \frac{1}{2}I(A:R)_\psi \,;\quad e^\infty = \frac{1}{2}I(A:B)_\psi.
\ee
\end{theorem}
\begin{proof}
To prove the above theorem we make use of the following relation (Theorem 1 of \cite{TCR09IT}): $\forall \rho^{AB}\in\cD(\cH_{A}\otimes\cH_B)$,
\begin{eqnarray}
\lim_{\eps \to 0}\lim_{n\to \infty}\frac{1}{n} H_{\rm min}^\eps(A|B)_{\rho^{\otimes n}}&=& H(A|B)_\rho \label{EQ_iid_min1}
\end{eqnarray}
and the following identity given by Lemma~\ref{LEM_IID}:
\begin{eqnarray}
\lim_{\eps \to 0}\lim_{n\to \infty}\frac{1}{n} H_{\rm 0}^\eps(A)_{\rho^{\otimes n}}&=& H(A)_\rho. \label{EQ_iid_min3}
\end{eqnarray}
In the above, $H(A)_\rho := - \tr \rho^A \log \rho^A$ is the von Neumann entropy of $\rho^A$, and $H(A|B)_\rho :=  H(AB)_\rho - H(A)_\rho$.

We first show that the upper and lower bounds for the quantum communication cost converge to the quantity $\frac{1}{2} I(A:R)_\psi$.  We have
\bea
{q^\infty}&:=& \lim_{\eps \to 0}\lim_{n\to \infty}\frac{1}{n}{{q}^{(1)}_{\eps, n}}\nonumber \\
&\le& \lim_{\eps \to 0}\lim_{n\to \infty}\frac{1}{n} \left(\frac{1}{2}\bigl[{H_{\rm 0}^{{\delta}}(A)_{\psi_n}}-{H_{\rm min}^{\delta}(A|R)_{\psi_n}}\bigr]-\log\delta'\right) \nonumber\\
&=&\frac{1}{2}\left[H(A)_\psi-H(A|R)_{\psi}\right] \nonumber \\
&=&\frac{1}{2}I(A:R)_{\psi}\label{upq01}.
\eea
The first line follows from the definition of $q^\infty$ in \reff{infq}. The second line follows from the upper bound for $q^{(1)}_{\eps,n}$ in Theorem~\ref{theorem5}. The third line follows from the identities \reff{EQ_iid_min1}-\reff{EQ_iid_min3} and the fact that $(\log\delta')/n$ term clearly vanishes as $n\to \infty$. Similarly, the lower bound for $q^{(1)}_{\eps,n}$ in Theorem~\ref{thmconverse} gives
\bea
{q^\infty}&:=& \lim_{\eps \to 0}\lim_{n\to \infty}\frac{1}{n}{{q}^{(1)}_{\eps, n}}\nonumber \\
&\geq& \lim_{\eps \to 0}\lim_{n\to \infty}\frac{1}{n} \left(\frac{1}{2}\bigl[{H_{\min}^{{\eps}}(A)_{\psi_n}}-{H_{\rm min}^{\overline{\eps}}(A|R)_{\psi_n}}\bigr]-\log\frac{\sqrt{2}}{\eps}\right) \nonumber\\
&=&\frac{1}{2}I(A:R)_{\psi}\label{upq02}.
\eea

Next, we can show that the upper and lower bounds for entanglement gain converge to the quantity $\frac{1}{2}I(A:B)_\psi$. Similarly, from  \reff{infe} and the lower bound for $e^{(1)}_{\eps,n}$ in Theorem~\ref{theorem5}, we have
\bea
{e^\infty}&:=& \lim_{\eps \to 0}\lim_{n\to \infty}\frac{1}{n}{{e}^{(1)}_{\eps, n}}\nonumber \\
&\ge& \lim_{\eps \to 0}\lim_{n\to \infty}\frac{1}{n} \left(\frac{1}{2}\bigl[{H_{\rm 0}^{{\delta}}(A)_{\psi_n}} + {H_{\rm min}^{\delta}(A|R)_{\psi_n}}\bigr] +\log\delta'\right) \nonumber\\
&=&\frac{1}{2}\left[H(A)_\psi + H(A|R)_{\psi}\right] \nonumber \\
&=&\frac{1}{2}I(A:B)_{\psi}\label{upq03}.
\eea
Finally, the upper bound for $e^{(1)}_{\eps,n}$ in Theorem~\ref{thmconverse} gives
\bea
{e^\infty}&:=& \lim_{\eps \to 0}\lim_{n\to \infty}\frac{1}{n}{{e}^{(1)}_{\eps, n}}\nonumber \\
&\leq& \lim_{\eps \to 0}\lim_{n\to \infty}\frac{1}{n} \left(\bigl[{q^{(1)}_{\eps,n} } + {H_{\rm min}^{\overline{\eps}}(A|R)_{\psi_n}}\bigr] +\log\frac{2}{\eps^2}\right) \nonumber\\
&=&\frac{1}{2} I(A:R)_\psi +H(A|R)_{\psi}\\
&=&\frac{1}{2}I(A:B)_{\psi}\label{upq}.
\eea
\end{proof}

%%%%%%%%%%%%%%%%%%%%%%%%%%%%%%%%%%%%%%%%%%%%%%%%%%%%%%%%%%
% Optimal rate of one-shot quantum communication
%%%%%%%%%%%%%%%%%%%%%%%%%%%%%%%%%%%%%%%%%%%%%%%%%%%%%%%%%%
\subsection{Optimal rate of one-shot quantum communication}
\label{qcap}
Our result (Theorem \ref{theorem5}) on the one-shot FQSW can be employed to provide a convenient characterization of the $\eps$-error one-shot quantum capacity of a noisy channel, entirely in terms of the smooth max-entropy. We consider this result to be an improvement over previously obtained results \cite{{BD10IT}} for reasons given in Section \ref{child}. Before stating our result we first describe the protocol of quantum communication that we are considering, and define the $\eps$-error one-shot quantum capacity.
\smallskip

\noindent
{\em{The protocol \cite{BD10IT}:}} Given a quantum channel $\cN^{A' \to B}$, let ${\cal{H}}_M$ be an $m$-dimensional subspace of its input Hilbert space $\cH_{A'}$, and let $0<\eps\leq 1$ be a fixed positive constant. Alice prepares a maximally entangled state $|\Phi^{A'A}_\cM\>\in \cM' \otimes \cM$, where $\cM' \simeq \cM\subset \cH_{A'}\simeq\cH_{A}$, and sends the part $A'$ through the channel $\cN^{A'\to B}$ to Bob. Bob is allowed to do any decoding operation (CPTP map) on the state that he receives. The final objective is for Alice and Bob to end up with a shared state which is nearly maximally entangled over $\cM' \otimes \cM$, its overlap with $|\Phi^{A'A}_\cM\>$ being at least $(1-\eps^2)$. There is no classical communication possible between Alice and Bob. In this scenario, the one-shot $\eps$-error quantum capacity of the channel $\cN^{A'\to B}$ is defined as follows:

\begin{definition}[One-shot $\eps$-error quantum capacity]
Given a quantum channel $\cN:{\cal D}(\sH_{A'})\mapsto{\cal D}(\sH_B)$ and a real number $0<\eps\leq 1$, the one-shot $\eps$-error quantum capacity of $\cN^{A'\to B}$ is defined as follows:
\begin{equation}\nonumber
Q^{(1)}_\eps(\cN):=\max\{\log m: \F_\ent(\cN;m)\ge 1-\eps^2\},
\end{equation}
where
\[
\F_\ent(\cN;m):=\max_{\cH_M\subseteq\cH_A\atop{\dim\cH_M=m}}\max_{\cD}\<\Phi^{A'A}_\cM|(\id\otimes\cD\circ\cN)(\Phi^{A'A}_\cM)|\Phi^{A'A}_\cM\>,
\]
with $\cD^{B\to A'}$ being a decoding CPTP-map.
\end{definition}

\begin{theorem}\label{one-qcap}
For any $0<\eps\leq 1$, the one-shot $\eps$-error quantum capacity of a noisy channel $\cN\equiv \cN^{A'\rightarrow B}$, satisfies the following bounds
\begin{equation}\label{q_bounds}
 \max_{\sS\subseteq\sH_A}\left[- H_{\max}^{\delta}(A|B)_{\psi_\sS}\right]
+ 2 \log {\delta'}
 \le \ Q^{(1)}_\eps(\cN)
\le \max_{\sS\subseteq\cH_A}\left[- H_{\max}^\eps(A|B)_{\psi_\sS}\right],
\end{equation}
for some $\delta>0$ such that $\eps:=2\sqrt{5\delta'}+2\sqrt{\delta}$ and $\delta'=\delta+\sqrt{4\sqrt{\delta}-4\delta}$, and ${\psi_\sS}$ denotes the state
\be\label{q_state}
|\psi_\sS\>^{ABE}:=(\bbI_A\otimes U^{A'\to BE}_{\cN})|\Phi^{AA'}_\sS\>.
\ee
In the above, $U_{\cN}^{A'\to BE}$ denotes a Stinespring isometry realizing the channel, and $|\Phi^{AA'}_\sS\>$ is a maximally entangled state of rank $m  (= {\rm{dim }} \sS)$ in $\cH_A\otimes\cH_{A'}$, $\sS \subseteq \cH_A$.
\end{theorem}
\begin{proof}
The lower bound in \reff{q_bounds} follows from the resource inequality \reff{q_RI} given in Section~\ref{child}. The upper bound in \reff{q_bounds} for the $\eps$-error one shot quantum capacity $Q^{(1)}_\eps(\cN)$ can be proved as follows. Denote by $\omega^{AA'}:=(\id_A\otimes\cD^{B\to A'})(\psi_\cM^{AB})$ Bob's decoded state, where $\psi_\cM^{AB}$ is the channel output defined through (\ref{q_state}). For any $\eps$-error one-shot quantum communication, we have
\be\label{q_cap_ball}
\Phi_\cM^{AA'}\in \cB^\eps(\omega^{AA'}).
\ee
Then an upper bound for $Q^{(1)}_\eps(\cN)$ can be obtained as follows.
\bea
\log m &=& - H_{\max} (A|A')_{\Phi_\cM} \nonumber \\
&\leq & \max_{\overline{\sigma}^{AA'}\in\cB^\eps(\omega^{AA'})} \left[-H_{\max}(A|A')_{\overline{\sigma}}\right] \nonumber \\
&=& - H_{\max}^\eps (A|A')_\omega  \nonumber \\
&\leq & - H_{\max}^\eps (A|B)_{\psi_\cM}.
\eea
The first inequality follows from (\ref{q_cap_ball}), and the second inequality follows from the data-processing inequality for $\eps$-smooth max-entropy (Lemma \ref{data}).
\end{proof}

It is interesting to compare Theorem~\ref{one-qcap} with Theorem~1 of \cite{BD10IT}. In the latter, the lower and upper bounds on 
the one-shot $\eps$-error quantum capacity were given in terms of {\em{different}} smoothed versions of the entropic quantity
$ H_0(A|B)_\rho$ defined by \reff{ent_H0}. The lower bound was given in terms of the quantity $[- \widetilde{H}_{0}^{\eps}(A|B)_{\psi_{\sS}}]$, where
\begin{equation}\label{EQ_smooth_H0}
\widetilde{H}_0^{\eps}(A|B)_\rho:=\min_{\overline{\rho}^{AB}\in\cB^{\eps}(\rho^{AB})} H_0(A|B)_{\overline{\rho}},
\end{equation}
is related to ${H}_{\max}^{\eps}(A|B)_\rho$ through Lemma~\ref{berta}, whereas the upper bound was given by an 
operator-smoothed version of $[-H_{0}(A|B)_{\psi_{\sS}}]$ (see \cite{BD10IT} for details). In contrast, our Theorem~\ref{one-qcap} has the advantage of being entirely given in terms of a single quantity, namely the $[-H_{\max}^{\eps}(A|B)_{\psi_{\sS}}]$.

We now consider the case of a memoryless quantum channel $\cN^{A'\to B}$.
Let us denote the output of $n$ successive, independent uses of the channel, corresponding to the input state $\ket{\Phi_{\cM_n}^{A'A}}\in \cM_n'\otimes \cM_n$, where $\cM_n\subset \cH_{A}^{\otimes n}$, by
\begin{equation}\label{psi_Mn}
\psi^{ABE}_{\cM_n}:=(\bbI_A\otimes U_{\cN}^{A'\to BE})^{\otimes n}(\Phi_{\cM_n}^{A'A}).
\end{equation}
Let $Q_{\eps}^{(1)}(\cN^{\otimes n})$ denote the one-shot $\eps$-error quantum capacity of the channel $\cN^{\otimes n}$. The quantum capacity $Q^\infty(\cN)$ of a memoryless channel $\cN$, which is evaluated in the limit of asymptotically many uses of the channel, can be defined in terms of $Q_{\eps}^{(1)}(\cN)$ as follows:
\[
Q^\infty(\cN) :=\lim_{\eps\to 0}\lim_{n\to\infty}\frac{1}{n}Q_{\eps}^{(1)}(\cN^{\otimes n}).
\]
Next we prove how the known expression \cite{Devetak03,BD10IT} of the quantum capacity $Q^\infty(\cN)$ can be recovered from Theorem~\ref{one-qcap}.
%%%%%%%%%%%%%%%%%%%%%%%%%
% Quantum Capacity Theorem
%$$$$$$$$$$$$$$$$$$$$$$$$$$$$$$$$$$$$
\begin{theorem}\label{q_asymp}
The quantum capacity of a memoryless channel $\cN$ is given by the following:
\be\label{q_arates}
Q^\infty(\cN) = \lim_{n\to\infty}\frac{1}{n}\max_{\cM_n\subseteq \cH_A^{\otimes n}}I(A\rangle B)_{\psi_{\cM_n}}
\ee
where $\psi^{ABE}_{\cM_n}$ is defined in \reff{psi_Mn}, and $I(A\rangle B)_{\psi_{\cM_n}}$ is the coherent information \reff{coherent} of the state $\psi_{\cM_n}$.
\end{theorem}
\begin{proof}
Note that
\begin{eqnarray}
Q^\infty(\cN) &:=& \lim_{\eps\to 0}\lim_{n\to\infty}\frac{1}{n}Q_{\eps}^{(1)}(\cN^{\otimes n})\nonumber \\
&\leq& \lim_{\eps\to 0}\lim_{n\to\infty}\frac{1}{n}\left( \max_{\cM_n\subseteq\cH_A^{\otimes n}}\left[- H_{\max}^{\eps}(A|B)_{\psi_{\sS_n}}\right] \right)\nonumber \\
&\leq& \lim_{\eps\to 0}\lim_{n\to\infty}\frac{1}{n}\left( \max_{\cM_n\subseteq\cH_A^{\otimes n}}\left[- H(A|B)_{\psi_{\sS_n}}+8\eps\log|A_{\cM_n}|+2h(2\eps)\right]\right)\nonumber \\
&=&\lim_{n\to\infty}\frac{1}{n}\left( \max_{\cM_n\subseteq\cH_A^{\otimes n}}\left[- H (A|B)_{\psi_{\sS_n}}\right]\right).
\end{eqnarray}
The first inequality follows from the upper bound in Theorem~\ref{one-qcap}. The second inequality follows from the fact that $|A_{\cM_n}|$ denotes the dimension of the Hilbert space on which the state $\psi_{\cM_n}^A$ is supported and from Lemma~5 in Ref.~\cite{RR10}, which states that for $0<\eps\leq 1$ and $\rho^{AB}\in\cD(\cH_A\otimes\cH_B)$
\[
H_{\rm max}^{\eps}(A|B)_\rho\geq H(A|B)_\rho - 8\eps\log|A|-2h(2\eps)
\]
where $h(\eps)=-\eps\log\eps-(1-\eps)\log(1-\eps)$.
 
Similarly, from the lower bound in Theorem~\ref{one-qcap} we have
\begin{eqnarray}
Q^\infty(\cN) &:=& \lim_{\eps\to 0}\lim_{n\to\infty}\frac{1}{n}Q_{\eps}^{(1)}(\cN^{\otimes n})\nonumber \\
&\geq& \lim_{\eps\to 0}\lim_{n\to\infty}\frac{1}{n}\left( \max_{\cM_n\subseteq\cH_A^{\otimes n}}\left[- H_{\max}^{\delta}(A|B)_{\psi_{\sS_n}}\right]+ 2 \log {\delta'} \right)\nonumber \\
&\geq& \lim_{\eps\to 0}\lim_{n\to\infty}\frac{1}{n}\left( \max_{\cM^{\otimes n}\subseteq\cH_A^{\otimes n}}\left[- H_{\max}^{\delta}(A|B)_{\psi_{\sS}^{\otimes n}}\right]+ 2 \log {\delta'} \right)\nonumber \\
&=& \max_{\cM\subseteq\cH_A}\left[- H (A|B)_{\psi_{\sS}}\right].
\end{eqnarray}
The second inequality follows since the maximization is over a smaller set. The last equality follows from 
the fact that $(\log\delta')/n$ vanishes as $n\to\infty$, and from the following relation (Theorem 1 of \cite{TCR09IT}): $\forall \rho^{AB}\in\cD(\cH_{A}\otimes\cH_B)$,
\begin{equation}
\lim_{\eps \to 0}\lim_{n\to \infty}\frac{1}{n} H_{\rm max}^\eps(A|B)_{\rho^{\otimes n}}= H(A|B)_\rho. \label{EQ_iid_max2}
\end{equation}
We can then obtain the lower bound for \reff{q_arates} by noting that $I(A\rangle B)_{\psi_{\cM_n}}=-H(A|B)_{\psi_{\cM_n}}$ and using the standard blocking argument.
\end{proof}

%%%%%%%%%%%%%%%%%%%%%%%%%%%%%%%%%%%%%%%%%%%%%%%%%%%%%%%%%
% Resource inequalities and the children protocols
%%%%%%%%%%%%%%%%%%%%%%%%%%%%%%%%%%%%%%%%%%%%%%%%%%%%%%%%%
\section{Resource inequalities and the children protocols}
\label{child}
As shown in \cite{DHW04PRL,ADW09PRSA}, the FQSW protocol in the asymptotic~i.i.d.~scenario can be conveniently expressed using a resource inequality \cite{DHW04PRL,DHW08IT,ADW09PRSA}. Before stating it, let us briefly recall the notation used in the resource inequality (RI) framework and how one interprets the inequalities.

Some of the basic units (referred to as {\em{unit}} asymptotic resources) of the RI framework are the following: $[c\rightarrow c]$ represents one bit of classical communication from Alice (the sender) to Bob  (the receiver); $[q\rightarrow q]$ represents one qubit of quantum communication from Alice  to Bob; $[qq]$ represents an ebit shared between Alice and Bob. For a more complete list of units, see \cite{DHW08IT}.

The original mother protocol (or quantum communication-assisted entanglement distillation), of which the FQSW is a generalization, is given by the following resource inequality \cite{DHW04PRL}:
\be\label{resource_mother}
\langle \psi^{AB}\rangle + \frac{1}{2} I(A:R)_\psi [q\rightarrow q] \ge
\frac{1}{2}I(A:B)_\psi[qq].
\ee
It states that $n$ copies of a state $\psi^{AB}$ shared between Alice and Bob (with purification $\psi^{ABR}$, where $R$ denotes the reference system) can be converted into $\frac{1}{2}I(A:B)_\psi$ EPR pairs per copy, under the condition that Alice is allowed to communicate with Bob by sending him qubits at a rate $\frac{1}{2}I(A:R)_\psi$ per copy. Minor inaccuracies in the final state are allowed, provided they vanish asymptotically, i.e., as $n \rightarrow \infty$. The one-shot analogue of the resource inequality \reff{resource_mother} is given by \reff{RI_mother-1} below.
In contrast, the RI for the FQSW protocol is given by \cite{ADW09PRSA}
\be\label{RI_FQSW}
\langle U^{S\rightarrow AB} : \psi^{S}\rangle + \frac{1}{2} I(A:R)_\psi [q\rightarrow q] \ge
\frac{1}{2}I(A:B)_\psi[qq] + \langle {\rm{id}}^{S\rightarrow B} : \psi^{S}\rangle.
\ee
In the above, $U^{S\rightarrow AB} $ is an isometry taking the system $S$ to $AB$ and signifies that a state is distributed between Alice and Bob, whereas the identity map ${\rm{id}}^{S\rightarrow B}$ on the right hand side signifies that the same state is given to Bob alone. Thus the inequality expresses that Alice transfers her part of the initial state to Bob. More precisely, it states that starting from the state $|\psi^{ABR}\rangle^{\otimes n}$, Alice can transfer her part of the state (and hence also the entanglement that she shares with the reference system $R$) to Bob by using quantum communication at the rate of $\frac{1}{2} I(A:R)_\psi $, and they can simultaneously distill EPR pairs at the rate $\frac{1}{2} I(A:B)_\psi$. Again, minor inaccuracies in the final state are allowed provided they vanish asymptotically. The one-shot version of this RI is given by \reff{RI_FQSW-1}. Note that resource inequalities provide statements about achievable rates of protocols, but do not ensure the optimality of these rates.

Next we state one-shot resource inequalities for the FQSW and some of its children protocols. A more exhaustive study of one-shot resource inequalities of all the different children protocols will be done in a later paper \cite{later}.

The RI for the {{one-shot FQSW}} for the state $|\psi\rangle^{ABR}$ is given by
\be\label{RI_FQSW-1}
\langle U^{S\rightarrow AB} : \psi^{S}\rangle + q^{(1)}_\eps [q\rightarrow q] \ge_\eps
e^{(1)}_\eps[qq] + \langle {\rm{id}}^{S\rightarrow B} : \psi^{S}\rangle,
\ee
where the quantum communication cost $ q^{(1)}_{\eps}$ and entanglement gain $e^{(1)}_{\eps}$ are given by
\begin{eqnarray}
q_{\eps}^{(1)}&=& \frac{1}{2}\left[H_{\rm 0}^{{\delta}}(A)_\psi-H_{\rm min}^{\delta}(A|R)_{\psi}\right]-\log\delta' \label{q1}\\
e_{\eps}^{(1)} &=& \frac{1}{2}\left[H_{\rm 0}^{{\delta}}(A)_\psi+H_{\rm min}^{\delta}(A|R)_\psi\right]+\log\delta',\label{e1}
\end{eqnarray}
for some $\delta>0$ such that $\eps=2\sqrt{5\delta'}+2\sqrt{\delta}$ and $\delta'=\delta+\sqrt{4\sqrt{\delta}-4\delta}$. This follows directly from Theorem \ref{theorem5}. Note that in \reff{RI_FQSW-1} the notation $\ge_\eps$ is used to denote that we are considering $\eps$-error one-shot FQSW.

The RI for the one-shot FQSW in \reff{RI_FQSW-1} yields the following RI for the one-shot state merging protocol when combined with the RI for teleportation, $2[c\rightarrow c] + [qq] \ge [q\rightarrow q]$ \cite{DHW08IT}:
\begin{eqnarray}
\fl
\langle U^{S\rightarrow AB} : \psi^{S}\rangle + q^{(1)}_\eps [q\rightarrow q] &+ &2q^{(1)}_\eps [c\to c] \ge_\eps
e^{(1)}_\eps[qq] + \langle {\rm{id}}^{S\rightarrow B} : \psi^{S}\rangle+ 2q^{(1)}_\eps [c\to c] , \nonumber \\
&\geq_\eps&(e^{(1)}_\eps-q^{(1)}_\eps)[qq] + \langle {\rm{id}}^{S\rightarrow B} : \psi^{S}\rangle+ q^{(1)}_\eps[q\to q].  \label{RI_SM}
\end{eqnarray}
In the RI \reff{RI_SM}, the $ q^{(1)}_{\eps}$ qubits of quantum communication which are employed at the start of the protocol are recovered at the end. Hence they play the role of a catalyst. Thus we can obtain the achievable \emph{entanglement gain} $(e^{(1)}_\eps-q^{(1)}_\eps)$ for the one-shot state merging protocol in terms of the $\delta$-smooth conditional min-entropy $H_{\min}^\delta(A|R)_\psi$, modulo additional $\eps$-dependent terms. Note that the entanglement gain $(e^{(1)}_{\eps}-q^{(1)}_{\eps})$ in \reff{RI_SM} can be also achieved using a one-shot state merging protocol {\em{without any catalyst}}, a result recently announced in \cite{Dupuis2010}. That the quantity $H_{\min}^\delta(A|R)_\psi$ is optimal for the one-shot $\eps$-error state merging protocol is further justified by a corresponding converse proof given in \cite{Dupuis2010}. This leads us to write one-shot resource inequalities, modulo the consideration of the catalyst, as follows (for which we replace the symbol $\geq_\eps$ by ${\widetilde{\geq}}_\eps$).\newline
{\bf{State Merging}}:
\be
\langle U^{S\rightarrow AB} : \psi^{S}\rangle + 2q^{(1)}_\eps [c\to c] \widetilde{\ge}_\eps (e^{(1)}_\eps-q^{(1)}_\eps)[qq] + \langle {\rm{id}}^{S\rightarrow B} : \psi^{S}\rangle.
\ee

The RI  \reff{RI_FQSW-1} for the one-shot FQSW  directly leads to the RI for the {{one-shot mother}} for the state $|\psi\rangle^{ABR}$ (if one focuses on the entanglement distillation alone and ignores the additional task of state transfer which is required in FQSW):\newline
{\bf{Mother}}:
\begin{equation}\label{RI_mother-1}
\langle \psi^{AB}\rangle + q^{(1)}_\eps[q\rightarrow q] \ge_\eps
e^{(1)}_\eps[qq].
\end{equation}

From \reff{RI_mother-1}, we can obtain the RI for {{one-shot entanglement distillation}} by combining it with the RI for teleportation as follows. \newline
{\bf{Entanglement Distillation}}:
\begin{eqnarray}\label{RI_distillation-1}
\langle\psi^{AB}\rangle + q^{(1)}_{\eps}[ q\to q] + 2q^{(1)}_{\eps}[c\to c] &\geq_\eps& e^{(1)}_{\eps}[qq] +2q^{(1)}_{\eps}[c\to c] \nonumber \\
&\geq_\eps&(e^{(1)}_{\eps}-q^{(1)}_{\eps})[qq] + q^{(1)}_{\eps}[q\to q]. \nonumber \\
\Rightarrow \langle\psi^{AB}\rangle + 2q^{(1)}_{\eps}[c\to c] &\widetilde{\geq}_\eps& (e^{(1)}_{\eps}-q^{(1)}_{\eps})[qq].
\label{cat}
\end{eqnarray}
The quantities $q^{(1)}_{\eps}$ and $e^{(1)}_{\eps}$ appearing in \reff{RI_mother-1} and \reff{RI_distillation-1} are given by \reff{q1} and \reff{e1} respectively. We thus obtain an achievable one-shot entanglement distillation rate in terms of the $\delta$-smooth max-entropy $[-H_{\max}^{\delta}(A|B)_\psi]$, modulo additional $\eps$-dependent terms. Note that in  \cite{BD10JMP} an expression for an achievable rate of the one-shot entanglement distillation, \emph{without any catalyst}, was obtained in terms of $\widetilde{H}_0^\eps(A|B)_\psi$, an entropic quantity defined through \reff{ent_Ho_smooth}. 

Two children protocols of the mother are {\em{noisy teleportation}} and {\em{noisy superdense coding}}, which as their names suggest, correspond to teleportation and superdense coding with a general entangled state $\psi^{AB}$. The one-shot resource inequalities for these protocols are easily
obtained from \reff{RI_mother-1} as follows: \newline
{\bf{Noisy Teleportation}}:
\begin{eqnarray}\label{tele}
\langle\psi^{AB}\rangle + q^{(1)}_{\eps}[ q\to q] + 2e^{(1)}_{\eps}[c\to c] &\geq_\eps& e^{(1)}_{\eps}[qq] +2e^{(1)}_{\eps}[c\to c] \nonumber \\
&\geq_\eps&e^{(1)}_{\eps}[q\to q] .\nonumber\\
\Rightarrow \langle\psi^{AB}\rangle+ 2e^{(1)}_{\eps}[c\to c] &{\widetilde{\geq}}_\eps& (e^{(1)}_{\eps}-q^{(1)}_{\eps})[q\to q].
\end{eqnarray}
{\bf{Noisy Superdense Coding}}:
\begin{eqnarray}
\langle\psi^{AB}\rangle + q^{(1)}_{\eps}[ q\to q] + e^{(1)}_{\eps}[q\to q] &\geq_\eps& e^{(1)}_{\eps}[qq] +e^{(1)}_{\eps}[q\to q]\nonumber \\
&\geq_\eps&2e^{(1)}_{\eps}[c\to c]. \nonumber\\
\Rightarrow \langle\psi^{AB}\rangle+ (q^{(1)}_{\eps}+e^{(1)}_{\eps})[q\to q]
&{\geq}_\eps& 2e^{(1)}_{\eps}[c\to c].\label{RI_super}
\end{eqnarray}
The second inequality in \reff{tele} follows from the RI for teleportation, whereas in obtaining the second inequality in \reff{RI_super} we have made use of the RI for superdense coding \cite{DHW08IT}:
\[
[q\rightarrow q] + [qq] \ge 2[c\rightarrow c].
\]

\begin{figure}[ht]
\subfigure[hang,raggedright][Applying the one-shot FQSW protocol to the channel output state $\psi^{ABE}$ of the father protocol results in Bob and the reference system $R$ sharing a state $\Phi^{RB_0}$ (the superscript $B_0$ denotes the $A_0$ system has been transferred to Bob) which is $\eps$-close to a maximally entangled state -- the desired outcome of the father protocol. However, the FQSW protocol requires acting on the reference system, which is inaccessible to Alice and Bob.]{
\def\svgwidth{\columnwidth}
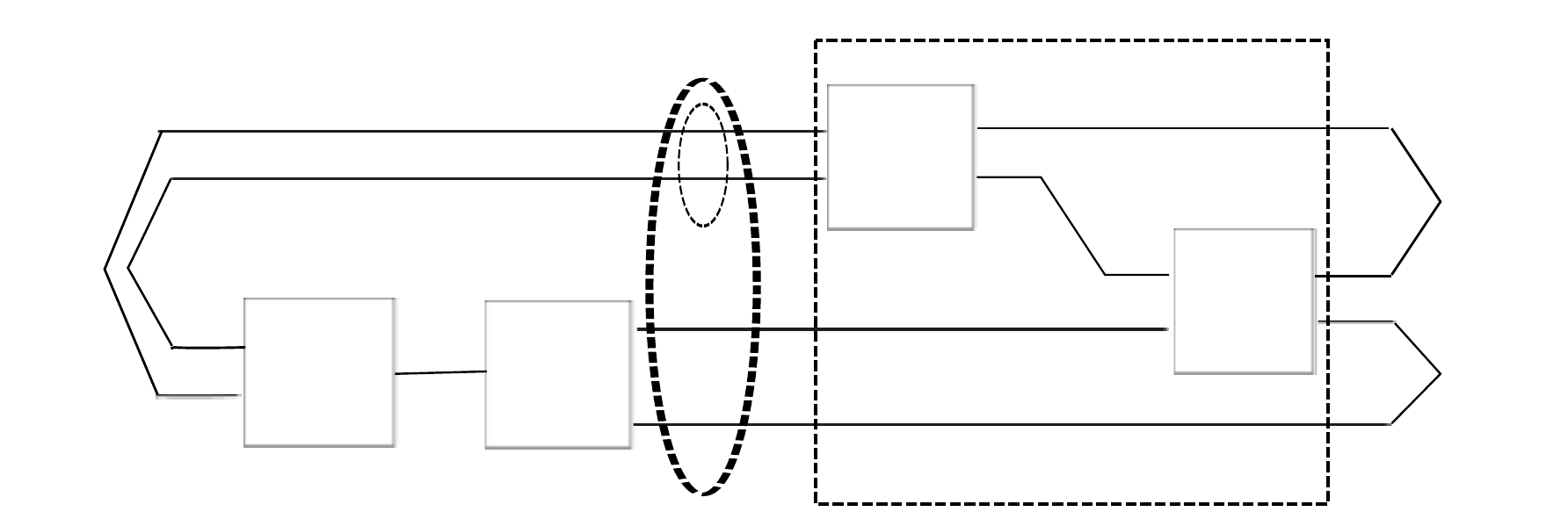
\label{fig:subfig1}
}
\subfigure[normal][Circumventing the problem of implementing the unitary $U$: Note that performing the unitary $U$ on system $A\equiv RB_1$ in the one-shot $\eps$-error FQSW protocol in \subref{fig:subfig1}  is equivalent to Alice performing $U^T$ on system $A_0A_1$ (since $RB_1$ and $A_0A_1$ are in a maximally entangled state) before employing her encoding isometry $W$.]
{
\def\svgwidth{\columnwidth}
 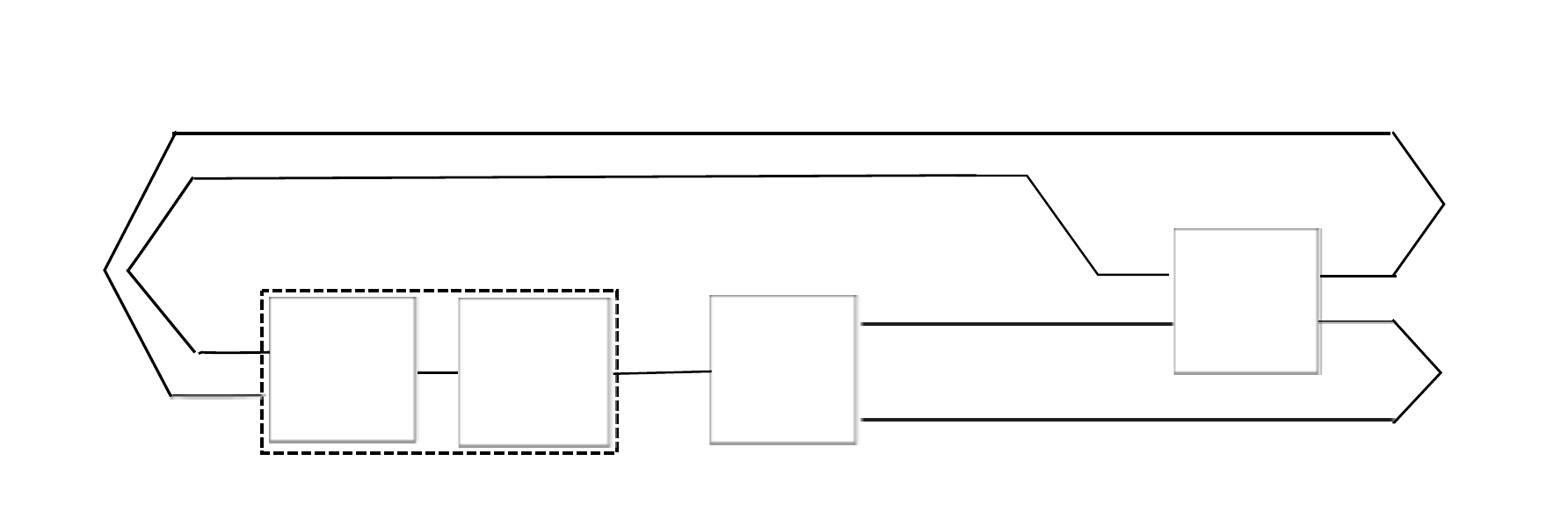 
\label{fig:subfig2}
}
\caption[Optional caption for list of figures]{Relating the one-shot FQSW and father protocols.}
\label{fig2}
\end{figure}

It has been proved in \cite{DHW04PRL,DHW08IT} that the entanglement gain and quantum communication cost for the FQSW protocol in the case in which asymptotically many copies of the tripartite state are shared between Alice, Bob and the reference, yield the entanglement cost and the quantum communication gain for the so-called {\em{father protocol}}, which is the protocol for entanglement-assisted quantum communication through a noisy channel. This is also the case in the one-shot regime, as explained below. Applying superdense coding after executing one instance of the father protocol allows us to trade quantum communication for classical communication and therefore results in one-shot entanglement-assisted classical communication through the noisy channel and yields a lower bound on the corresponding capacity. This mimics the method of determining an achievable rate for entanglement-assisted classical communication in the asymptotic scenario within the resource inequality framework \cite{DHW04PRL,DHW08IT}. As is natural in the one-shot scenario, we allow for a finite error (say $\eps$) for both the FQSW and father protocols.

The resource inequality for the one-shot FQSW yields a resource inequality for the following entanglement-assisted quantum communication (or father) protocol. Alice initially shares two maximally entangled states -- one ($\Phi^{A_1B_1}$) with Bob, and the other ($\Phi^{A_0R}$) with a reference system $R$ which is inaccessible to both her and Bob. The maximally entangled state $\Phi^{A_1B_1}$ acts as a resource of prior shared entanglement between Alice and Bob. Alice's goal is to send the quantum system $A_0$ to Bob through a noisy quantum channel $\cN^{A'\to B}$ so that finally Bob shares a maximally entangled state with the reference system. To achieve this goal she does an encoding isometry $W$ on the systems $A_0$ and $A_1$ in her possession (see Fig.~\ref{fig:subfig1}). The output $A'$ of this isometry is subjected to the Stinespring isometry $U_{\cN}^{A'\to BE}$ realizing the noisy channel $\cN^{A'\to B}$. Let us denote the resulting pure state by $\psi^{ABE}$, where $A\equiv RB_1$.

In order to see how one can relate the one-shot father protocol to the one-shot FQSW protocol, let us consider the state $\psi^{ABE}$ of the one-shot father protocol to be the initial tripartite state of an $\eps$-error one-shot FQSW protocol.

The one-shot FQSW theorem (Theorem~\ref{theorem5}) tells us that there exists a unitary operator $U^{A \to B_1R}$ and an isometry $V^{B_1B \to B_0B'}$, such that the state resulting from their successive actions, i.e., the state 
\[
(V^{B_1B \to B_0B'} \circ U^{A \to B_1R} )\psi^{ABE}(V^{B_1B \to B_0B'} \circ U^{A \to B_1R}) ^\dagger,
\] 
is $\eps$-close to the state
\[
\Phi^{B_0R} \otimes \psi^{B'E},
\]
with the systems $B_0$ and $B'$ being in Bob's possession, and $\psi^{B'E}$ being a purification of the state $\psi^E := \tr_{AB}\psi^{ABE}$ \footnote{Here the system $B'$ denotes the composite systems $AB$ of the state $\psi^{ABE}$. Furthermore, one can think of the state $\psi^{B'E}$ to be identical to the state $\psi^{ABE}$.}. The protocol corresponding to this theorem consists of performing the unitary $U^{A\to B_1R}$ on the system $A$ and sending the system $B_1$ to Bob, who then performs the isometry $V^{B_1B\to B_0B'}$ on the composite system $BB_1$. The number of qubits sent in the protocol is hence equal to $\log |B_1|$ and the number of ebits distilled is $\log|A_0|=\log |B_0|$ \footnote{Note that in the setting of Fig.~\ref{fig:subfig1}, sending the quantum register $B_1$ to Bob is not necessary since it is already in Bob's possession.}. Note that at the end of the protocol, Bob shares a state with the reference system $R$ which is $\eps$-close to a maximally entangled state and thus by using the $\eps$-error one-shot FQSW protocol on the state $\psi^{ABE}$ of the father protocol, the aim of the father protocol is achieved. Effectively, $\log |A_0|$ qubits have been transmitted from Alice to Bob and this therefore quantifies the quantum communication gain of the protocol. 

Note however that there is a caveat in the above argument. Since $A\equiv B_1R$, the unitary $U \equiv U^{A \to B_1R}$ requires acting on the reference system $R$. However, $R$ is inaccessible to Alice and Bob and hence the unitary cannot be implemented. This problem is easily overcome by noticing that performing the unitary $U$ on system $A\equiv RB_1$ in the one-shot $\eps$-error FQSW protocol in Fig.~\ref{fig:subfig1} is equivalent to Alice performing $U^T$ on system $A_0A_1$ since $RB_1$ and $A_0A_1$ are in a maximally entangled state. In this way, a complete one-shot $\eps$-error entanglement-assisted quantum communication protocol with encoding $W\circ U^T$ and decoding $V$ is established (see Fig.~\ref{fig:subfig2}). Note that the entanglement cost of this protocol (i.e., the required prior shared entanglement between Alice and Bob) is $\log|B_1|$ ebits, which is equal to the quantum communication cost of the one-shot FQSW protocol, and the quantum communication gain of the one-shot father protocol is equal to $\log |A_0|$, which in turn is equal to the entanglement gain of the one-shot FQSW. These considerations yield the following resource inequality for the one-shot $\eps$-error father protocol through a noisy quantum channel $\cN^{A'\to B}$: \newline
{\bf{Father}}:
\begin{equation}\label{1-father}
\langle\cN\rangle + {\widetilde{e}}^{(1)}_\eps[ qq] \geq_\eps {\widetilde{q}}^{(1)}_\eps [q\to q]
\end{equation}
where 
\begin{eqnarray}
{\tilde{e}}^{(1)}_\eps&=& \frac{1}{2}\left[H_{\rm 0}^{{\delta}}(A)_\psi-H_{\rm min}^{\delta}(A|E)_{\psi}\right]-\log\delta' \label{q11}\\
{\tilde{q}}^{(1)}_\eps &=& \frac{1}{2}\left[H_{\rm 0}^{{\delta}}(A)_\psi+H_{\rm min}^{\delta}(A|E)_\psi\right]+\log\delta'.\label{e11}
\end{eqnarray}
Here $\psi$ denotes the state $\ket{\psi}^{ABE}=U_{\cN}^{A'\to BE}\ket{\varphi}^{AA'}$, with $U_{\cN}^{A'\to BE}$
being a Stinespring isometry realizing the channel, and $\ket{\varphi}^{AA'}$ being some pure state in $\cH_A \otimes \cH_{A'}$.

The RI \reff{1-father} for the one-shot father protocol readily yields a RI for the one-shot entanglement-assisted classical communication through a noisy quantum channel, which in turn yields a lower bound on the one-shot entanglement-assisted classical capacity. This can be seen as follows. Combining \reff{1-father} with the resource inequality for superdense coding, yields the following resource inequality for one-shot entanglement-assisted classical communication through a noisy channel $\cN\equiv \cN^{A'\rightarrow B}$: \newline
{\bf{Entanglement-assisted Classical Communication}}:
\begin{eqnarray}
\langle\cN\rangle + \widetilde{q}^{(1)}_{\eps}[qq] + \widetilde{e}^{(1)}_{\eps}[qq] &\geq_\eps& \widetilde{q}^{(1)}_{\eps}[q\to q]+\widetilde{q}^{(1)}_{\eps}[qq]\nonumber \\
&\geq_\eps& 2 \widetilde{q}^{(1)}_{\eps}[c\to c]. \nonumber\\
\Rightarrow \langle\cN\rangle+ (\widetilde{q}^{(1)}_{\eps}+\widetilde{e}^{(1)}_{\eps})[q q]
&\,\,{\geq}_\eps &\,\, 2\widetilde{q}^{(1)}_{\eps}[c\to c].\label{RI_super2}
\end{eqnarray}

 Combining \reff{1-father} with the trivial RI $[q\to q]\ge [qq]$, we can also obtain the RI for the one-shot quantum communication through a noisy channel $\cN\equiv \cN^{A'\rightarrow B}$: \newline
{\bf{Quantum Communication}}:
\begin{eqnarray}\label{q_RI}
\langle\cN\rangle & {\widetilde{\geq}}_\eps & Q^{(1)}_\eps [q\to q]
\end{eqnarray}
where
\bea
Q^{(1)}_\eps &:=&  {\tilde{q}}^{(1)}_\eps-  {\tilde{e}}^{(1)}_\eps \nonumber\\
&=& H_{\rm min}^{\delta}(A|E)_{\psi}+2\log \delta' \nonumber\\
&=& - H_{\rm max}^{\delta}(A|B)_{\psi}+2\log \delta'. \label{newq}
\eea
In \reff{newq} the last equality follows from the duality relation \reff{EQ_dual} between the min- and max-entropy. By using a decoupling theorem analogous to Theorem \ref{thm_decoupling}, we can prove that $Q^{(1)}_\eps$ is indeed an achievable rate for the one-shot $\eps$-error quantum communication through the channel $\cN$, and this leads to the lower bound in Theorem \ref{one-qcap}.

Another important protocol, of which the FQSW is a primitive, is state redistribution. In its one-shot version, the protocol is as follows. Alice and Bob share a tripartite state $\rho^{ABC}$, where Alice holds systems $A$ and $C$, and Bob holds system $B$. Let the state $\rho^{ABC}$ be purified by a reference system $R$, the pure state being denoted as $\ket{\psi}^{ABCR}$. The task is for Alice to transfer her system $A$ to Bob while keeping the overall purification $\ket{\psi}^{ABCR}$ unchanged (possibly with the help of prior shared entanglement). Our results on the one-shot FQSW imply that Alice can achieve this task (up to a finite accuracy $(1-\eps)$)
by sending
$$
\frac{1}{2}I_{\max}^\eps(A:R|B )_\psi:= \frac{1}{2}\left[H_{\rm max}^\eps(A|B)_{\psi}-H_{\rm min}^\eps(A|RB)_{\psi}\right]
$$
number of qubits to Bob, modulo an additional $\eps$-dependent factor. The proof of this statement and its optimality will be presented in a forthcoming paper \cite{jonathan}.

%%%%%%%%%%%%%%%%%%%%%%%%%%%%%%%%%%%%%%%%%%%%%%%%%%%
% Proof
%%%%%%%%%%%%%%%%%%%%%%%%%%%%%%%%%%%%%%%%%%%%%%%%%%%
\section{Proofs of Theorem~\ref{theorem5} and Theorem~\ref{thmconverse}}%
\label{SEC_PROOF}%
\subsection{Achievability proof of the one-shot FQSW protocol}\label{achieve}

The proof employs the following one-shot decoupling theorem which is proved in Appendix~\ref{Appendix_B} for completeness. Various versions of the proofs can be found in, e.g., \cite{BD10IT,BCR09,DupuisThesis,Dupuis2010}.

\begin{theorem}[One-shot decoupling]\label{thm_decoupling}
Fix $0<\eps\leq 1$, and $\rho^{AR}\in\cD(\cH_{A}\otimes\cH_R)$. Let $A_1A_2$ be a decomposition of the system $A$.
Define
\begin{equation}\label{EQ_dir_one-shot00}
\sigma^{A_1R}(U)=\tr_{A_2} \left[(U\otimes\bbI_R)\rho^{AR}(U\otimes\bbI_R)^\dagger \right],
\end{equation}
where $U$ be a unitary acting on system $A$. If
\begin{equation}\label{EQ_dir_one-shot01}
\log |A_1| \leq \frac{1}{2}\Bigl[\log|A|+H_{\rm min}^\eps(A|R)_\rho\Bigr]+\log\eps
\end{equation}
then
\begin{equation}\label{EQ_dir_one-shot02}
\int_{U(A)}\|\sigma^{A_1R}(U)-\tau^{A_1}\otimes\rho^R\|_1\ dU\leq 5\eps,
\end{equation}
where $dU$ is the Haar measure over the unitaries on system $A$.
\end{theorem}

\begin{proof}{\bf{of Theorem \ref{theorem5}:}} Fix $0<\eps\leq 1$ and a pure state $\ket{\psi}^{ABR}$. For any $0<\delta\leq 1$ to be specified later, let $Q$ be the operator for which the minimum in the definition
\begin{equation}\label{size_Q}
H_0^{{\delta}}(A)_\psi:=\min_{0\leq Q\leq \bbI_A\atop{\tr Q\psi^A\geq1-\delta}}\log\tr\Pi_{\sqrt{Q}\psi^A\sqrt{Q}}
\end{equation}
is achieved. Define
\[
\psi_Q^{ABR}:=(\sqrt{Q}\otimes\bbI_{BR})\psi^{ABR}(\sqrt{Q}\otimes\bbI_{BR}),
\]
and let $\psi_Q^{AR},\psi_Q^A$ denote its reduced states.
It follows from the Gentle Measurement lemma, Lemma~\ref{gentle}, that 
\[
\|\psi_Q^{ABR}-\psi^{ABR}\|\leq 2\sqrt{\delta},
\]
since $\tr[(Q\otimes\bbI_{BR})\psi^{ABR}]\geq 1-\delta$. 

Denote the dimension of Hilbert space on which the state $\psi_Q^A$ is supported by $|A_Q|$.  Due to the choice of $Q$ in \reff{size_Q}, we have
\[
\log|A_Q|:=\log\tr\Pi_{\psi_Q^A}=H_0^\delta(A)_\psi,
\]
where $\Pi_{\psi_Q^A}$ denotes the projector onto the support of $\psi^A_Q$.
Applying the one-shot decoupling theorem (Theorem~\ref{thm_decoupling}) to the state $\psi_Q^{AR}$ shows that, for any $\delta'\equiv\delta+\sqrt{4\sqrt{\delta}-4\delta}$
\[
\log |A_1| \leq \frac{1}{2}\left[\log|A_Q|+H_{\rm min}^{\delta'}(A|R)_{\psi_Q}\right]+\log\delta',
\]
there exists an isometry $U^{A\to A_1A_2}$ such that if Alice acts on her share of the tripartite pure state $\psi_Q^{ABR}$ with it and sends the system $A_2$ to Bob, then the state of the system $A_1$, which she retains, is decoupled from the state of the reference system
$R$, i.e.,
\[
\|\Omega_Q^{A_1R}-\tau^{A_1}\otimes\psi_Q^R\|_1\leq 5 \delta',
\]
where $\Omega_Q^{A_1R}$ is the reduced density matrix of the following pure state
\[
\ket{\psi_Q}^{A_1A_2BR}=U^{A\to A_1A_2}\ket{\psi_Q}^{ABR}.
\]
Note that to send the system $A_2$ to Bob, Alice needs to transmit $\log|A_2|$ qubits to Bob. We denote the resulting pure state by $\ket{\Omega_Q}^{A_1B_2BR}$, where we have replaced $A_2$ by $B_2$ since it is now in Bob's possession.

Note that since $\Phi^{A_1B_1}\otimes\psi_Q^{ABR}$ is a purification of the state $\tau^{A_1}\otimes\psi_Q^R$, and all purifications are related by isometries, it follows from Uhlmann's theorem \cite{uhlmann} that there must exist an isometry $\cV^{B_2B\to B_1B'B}$, with ${\cH}_{B_1} \simeq {\cH}_{A_1}$, which Bob can employ on the systems $B_2$ and $B$ now in his possession to obtain the following decoded state
\[
\ket{\widehat{\Omega}_Q}^{A_1B_1B'BR}=\cV^{BB_2\to B_1B'B}\ket{\Omega_Q}^{A_1B_2BR}
\]
such that it is $\eps$-close to the optimal state $\Phi^{A_1B_1}\otimes\psi^{B'BR} $:
\begin{eqnarray}
\fl
\|\widehat{\Omega}_Q^{A_1B_1B'BR}-\Phi^{A_1B_1}\otimes\psi^{B'BR} \| \leq\|\widehat{\Omega}_Q^{A_1B_1B'BR}-\Phi^{A_1B_1}\otimes\psi_Q^{B'BR} \| \nonumber \\
+\|\Phi^{A_1B_1}\otimes\psi_Q^{B'BR}-\Phi^{A_1B_1}\otimes\psi^{B'BR} \| \nonumber \\
\leq 2\sqrt{5\delta'}+2\sqrt{\delta}:=\eps.
\end{eqnarray}
Note that since the systems $A_1$ and $B_1$ are in a maximally entangled state, the protocol results in the generation of $e^{(1)}_{\eps}$ ebits of entanglement, where
\begin{eqnarray}
e^{(1)}_{\eps}&:=&\log|A_1| = \frac{1}{2}\left[\log|A_Q|+H_{\rm min}^{\delta'}(A|R)_{\psi_Q}\right]+\log\delta' \nonumber\\
&\geq& \frac{1}{2}\left[H^{{\delta}}_0(A)_\psi+H_{\rm min}^{\delta}(A|R)_{\psi}\right]+\log\delta'.
\end{eqnarray}
The above inequality follows because $\cB^{\delta}(\psi^{AR})\subset\cB^{\delta'}(\psi_Q^{AR})$. Further, the number of qubits that Alice needs to transmit to Bob is given by
\begin{eqnarray}
q^{(1)}_{\eps}&:=&\log|A_2|= \log |A_Q|- \frac{1}{2}\left[\log|A_Q|+H_{\rm min}^{\delta'}(A|R)_{\psi_Q}\right]-\log\delta' \nonumber\\
&\leq & \frac{1}{2}[H_{\rm 0}^{{\delta}}(A)_\psi- H_{\rm min}^{\delta}(A|R)_\psi]-\log\delta'.
\end{eqnarray}
\end{proof}

%%%%%%%%%%%%%%%%%%%%%%%%%%%%%%%%%%%%%%%
% Converse proof
%%%%%%%%%%%%%%%%%%%%%%%%%%%%%%%%%%%%%%%
\subsection{Converse proof of the One-shot FQSW protocol}
\label{converse}

\begin{proof}{\bf{of Theorem \ref{thmconverse}}}

Without loss of generality, {\em{any}} $\eps$-error FQSW protocol for a tripartite pure state
${\psi}^{ABR}$ can be described by a pair of encoding and decoding operations
(i.e., CPTP maps) $(\cE,\cD)$ as follows:
\begin{enumerate}
\item  Alice's encoding operation $\cE^{A\to A_1A_2}$ on the state $\psi^A=\tr_{BR}{\psi}^{ABR}$. Denote the state after Alice's operation by
    \begin{equation}\label{EQ_conv-state}
    \ket{\Omega}^{A_1A_2E_1BR}=U_{\cE}^{A\to A_1A_2E_1}\ket{\psi}^{ABR}
    \end{equation}
    where $U_{\cE}^{A\to A_1A_2E_1}$ is a Stinespring isometry of the map $\cE^{A\to A_1A_2}$. Alice then sends the system $A_2$ to Bob. This results in the
state $\ket{\Omega}^{A_1B_2E_1BR}$, where we have replaced $A_2$ by $B_2$ since
it is now in Bob's possession.
\item Bob's decoding map $\cD^{B_2 B\to B_1 B'B}$. Denote Bob's output state by $\widehat{\Omega}^{A_1B_1B'BR}:=\cD^{B_2 B\to B_1 B'B}(\Omega^{A_1B_2BR})$, where $\Omega^{A_1B_2BR}:= \tr_{E_1}{\Omega}^{A_1B_2E_1BR}$ and $\widehat{\Omega}^{A_1B_1B'BR}$ is the reduced density matrix of the following pure state
    \begin{equation}\label{EQ_conv-state02}
    \ket{\widehat{\Omega}}^{A_1B_1B'BRE_1E_2}=U_{\cD}^{B_2 B\to B_1B'BE_2}\ket{\Omega}^{A_1B_2E_1BR},
    \end{equation}
   with $U_{\cD}^{B_2 B\to B_1B'BE_2}$ being a Stinespring isometry of $\cD^{B_2 B\to B_1 B'B}$, such that
\begin{equation}\label{EQ_conv01}
\|\Phi^{A_1B_1}\otimes\psi^{B'BR}-\widehat{\Omega}^{A_1B_1B'BR}\|_1\leq \eps,
\end{equation}
where $\ket{\Phi}^{A_1B_1}$ is the maximally entangled state on systems $A_1B_1$.
\end{enumerate}
By Uhlmann's theorem \cite{uhlmann}, there exists a pure state $\varphi^{E_1E_2}\in {\cD}({\cH}_{E_1}\otimes\cH_{E_2}) $ such that
\begin{equation}\label{EQ_conv011}
\|\Phi^{A_1B_1}\otimes \psi^{B'BR}\otimes \varphi^{E_1E_2}- \widehat{\Omega}^{A_1B_1B'BRE_1E_2}\|_1\leq 2 \sqrt \eps.
\end{equation}
%%%%%%%%%%%%%%%%%%%
Combining (\ref{EQ_conv011}) with the monotonicity of the trace distance under
partial trace, we have
\begin{equation}\label{EQ_conv020}
\|\tau^{A_1}\otimes\psi^{R}\otimes\varphi^{E_1}-\widehat{\Omega}^{A_1RE_1}\|_1
\leq 2 \sqrt{\eps},
\end{equation}
where $\tau^{A_1}=\tr_{B_1}\Phi^{A_1B_1}$ is the completely mixed state on system $A_1$, $\psi^R=\tr_{B'B}\psi^{B'BR}$, and $\varphi^{E_1}=\tr_{E_2}\varphi^{E_1E_2}$. Note that $\Omega^{A_1RE_1}=\widehat{\Omega}^{A_1RE_1}$
because Bob's decoding operation $\cD^{B_2 B\to B_1 B'B}$ (or its corresponding
isometry $U_{\cD}^{B_2 B\to B_1B'BE_2}$ which relates the pure states
$\ket{\Omega}^{A_1B_2E_1BR}$ and $\ket{\widehat{\Omega}}^{A_1B_1B'BRE_1E_2}$)
does not affect the systems $A_1$, $R$ and $E_1$. Hence we can rewrite (\ref{EQ_conv020}) as
\begin{equation}\label{EQ_conv02}
\|\tau^{A_1}\otimes\psi^{R}\otimes\varphi^{E_1}-\Omega^{A_1RE_1}\|_1\leq 2 \sqrt \eps.
\end{equation}
%In the following, all the min- and max-entropies are defined with respect to the state in (\ref{EQ_conv-state}).

We make use of the following lemma:
\begin{lemma}\label{LEM_Dec} Fix $\delta\geq0$, and let $\kappa:=2\sqrt{\eps}$, then
\begin{equation}
H_{\min}^{\delta+2\sqrt{\kappa}}(A_1E_1|R)_\Omega \geq H_{\min}^\delta(A_1E_1)_\Omega
\end{equation}
	where $\Omega^{A_1E_1R}$ is the reduced density matrix of
the pure state defined in (\ref{EQ_conv-state}).
\end{lemma}
\begin{proof}
Using \reff{CC}, we infer from (\ref{EQ_conv02}) that
\be\label{29}
C(\tau^{A_1}\otimes\psi^{R}\otimes\varphi^{E_1},\Omega^{A_1RE_1})\leq
\sqrt{\kappa},
\ee
{{where}} $C(\rho, \sigma)$ is defined by \reff{crho} for any $\rho \in \cD(\cH)$ and $\sigma \in \cD_{\leq}(\cH)$.
Furthermore, the monotonicity property \reff{mono} under partial trace yields
\begin{equation}
C(\tau^{A_1}\otimes\varphi^{E_1},\Omega^{A_1E_1})\leq C(\tau^{A_1}\otimes\psi^{R}\otimes\varphi^{E_1},\Omega^{A_1RE_1})\leq \sqrt{\kappa}.
\end{equation}
Then for any $\overline{\omega}^{A_1E_1}\in \cB^{\delta}(\Omega^{A_1E_1})$
\[
C(\overline{\omega}^{A_1E_1},\tau^{A_1}\otimes\varphi^{E_1})\leq  C(\overline{\omega}^{A_1E_1},\Omega^{A_1E_1})+C(\tau^{A_1}\otimes\varphi^{E_1},\Omega^{A_1E_1})\leq \delta + \sqrt{\kappa},
\]
where the first inequality follows from the fact that $C(\rho, \sigma)$ is a metric
and hence satisfies the triangle inequality. Further,
\be\label{num}
C(\overline{\omega}^{A_1E_1}\otimes\psi^R,\tau^{A_1}\otimes\varphi^{E_1}\otimes\psi^R) =C(\overline{\omega}^{A_1E_1},\tau^{A_1}\otimes\varphi^{E_1})\leq \delta+ \sqrt{\kappa}.
\ee
Applying the triangle inequality once again and using \reff{29} and \reff{num} yield
\begin{eqnarray*}
\fl
C(\overline{\omega}^{A_1E_1}\otimes\psi^R,\Omega^{A_1E_1R})&\leq& C(\overline{\omega}^{A_1E_1}\otimes\psi^R,\tau^{A_1}\otimes\varphi^{E_1}\otimes\psi^R)+C(\tau^{A_1}\otimes\varphi^{E_1}\otimes\psi^R,\Omega^{A_1E_1R})\\
&\leq& \delta + 2\sqrt{\kappa}
\end{eqnarray*}
In other words,  $\forall \, \overline{\omega}^{A_1E_1}\in \cB^{\delta}(\Omega^{A_1E_1})$, the following is true:
\[
\overline{\omega}^{A_1E_1}\otimes\psi^R \in \cB^{\delta+2\sqrt{\kappa}}(\Omega^{A_1E_1R}).
\]
We then have
\begin{eqnarray}
\fl
H_{\rm min}^{\delta + 2\sqrt{\kappa}}(A_1E_1|R)_\Omega
&=& \max_{\overline{\sigma}^{A_1E_1R}\in \cB^{\delta+ 2\sqrt{\kappa}}(\Omega^{A_1E_1R})} H_{\rm min}(A_1E_1|R)_{\overline{\sigma}}\nonumber \\
&\geq& \max_{\overline{\omega}^{A_1E_1}\in \cB^{\delta}(\Omega^{A_1E_1})} H_{\rm min}(A_1E_1|R)_{\overline{\omega}^{A_1E_1}\otimes\psi^R}\nonumber \\
&=&\max_{\overline{\omega}^{A_1E_1}\in \cB^{\delta}(\Omega^{A_1E_1})}
\max_{\varrho^R\in\cD(\cH_R)}\left[
-D_{\max}(\overline{\omega}^{A_1E_1}\otimes\psi^R ||\bbI_{A_1E_1} \otimes \varrho^R)\right]\nonumber\\
&\geq&\max_{\overline{\omega}^{A_1E_1}\in \cB^{\delta}(\Omega^{A_1E_1})}
\left[
-D_{\max}(\overline{\omega}^{A_1E_1}\otimes\psi^R ||\bbI_{A_1E_1} \otimes
\psi^R)\right]\nonumber\\
&=&\max_{\overline{\omega}^{A_1E_1}\in \cB^{\delta}(\Omega^{A_1E_1})}
H_{\rm min}(A_1E_1)_{\overline{\omega}^{A_1E_1}} \nonumber \\
&=& H^{\delta}_{\rm min}(A_1E_1)_{\Omega}.
\end{eqnarray}
The first inequality follows because $\overline{\omega}^{A_1E_1}\otimes\psi^R \in \cB^{\delta+2\sqrt{\kappa}}(\Omega^{A_1RE_1})$ for any $\overline{\omega}^{A_1E_1}\in \cB^{\delta}(\Omega^{A_1E_1})$. The second inequality follows because we choose a particular $\varrho^R=\psi^R$.
\end{proof}

We can now obtain a lower bound for the one-shot quantum communication cost,
$q^{(1)}_\eps$, as follows. Let $\kappa:=2\sqrt{\eps}$. Then, for any
$\eps' >0$ and $\eps'',\eps'''\geq0$, we have
\begin{eqnarray}\label{67}
q^{(1)}_\eps &=& \log |A_2| \nonumber \\
&\geq & H_{\max}^{\eps''}(A_2|B)_\Omega \nonumber \\
&\geq& H_{\max}^{\eps'+2\eps''+\eps'''+2\sqrt{\kappa}}(A_2A_1E_1|B)_\Omega - H_{\max}^{\eps'''+2\sqrt{\kappa}}(A_1E_1|A_2B)_\Omega-\log\frac{2}{\eps'^2}\nonumber\\
&\geq& H_{\max}^{\eps'+2\eps''+\eps'''+2\sqrt{\kappa}}(A|B)_\psi - H_{\max}^{\eps'''}(A_1E_1|A_2BR)_\Omega-\log\frac{2}{\eps'^2}\nonumber\\
&\geq& H_{\max}^{\eps'+2\eps''+\eps'''+2\sqrt{\kappa}}(A|B)_\psi - H_{\max}^{\eps'''}(A_1A_2E_1|BR)_\Omega-\log|A_2|-\log\frac{2}{\eps'^2}\nonumber\\
&=& -H_{\min}^{\eps'+2\eps''+\eps'''+2\sqrt{\kappa}}(A|R)_\psi - H_{\max}^{\eps'''}(A|BR)_\psi-\log|A_2|-\log\frac{2}{\eps'^2}\nonumber\\
&=& -H_{\min}^{\eps'+2\eps''+\eps'''+2\sqrt{\kappa}}(A|R)_\psi + H_{\min}^{\eps'''}(A)_\psi-\log|A_2|-\log\frac{2}{\eps'^2}.
\end{eqnarray}
The first inequality follows from Lemma 20 of \cite{TCR10IEEEIT}. The second inequality follows from the chain rule for smooth max-entropy (Lemma \ref{chain}). The third inequality follows from the fact that, for any $\delta\geq0$, $H_{\rm max}^{\delta}(A|R)_\psi=H_{\rm max}^{\delta}(A_1A_2E_1|R)_\Omega$ since the two states $\ket{\psi}^{ABR}$ and $\ket{\Omega}^{A_1A_2E_1BR}$ are related by an isometry $U^{A\to A_1A_2E_1}_\cE$ (Lemma \ref{mincond}), and by a simple application of the duality relation (\ref{EQ_dual})
and Lemma~\ref{LEM_Dec}, which yields
\begin{eqnarray*}
- H_{\max}^{\eps'''+2\sqrt{\kappa}}(A_1E_1|A_2B)_\Omega &=& H_{\min}^{\eps'''+2\sqrt{\kappa}}(A_1E_1|R)_\Omega \\
&\geq& H_{\min}^{\eps'''}(A_1E_1)_\Omega \\
&=& - H_{\max}^{\eps'''}(A_1E_1|A_2BR)_\Omega.
\end{eqnarray*}
The fourth inequality of \reff{67} follows from Lemma~\ref{LEM_DIM}. The second equality follows from the duality relation (\ref{EQ_dual}) and the fact that, for any $\delta\geq0$, $H_{\rm max}^{\delta}(A|BR)_\psi=H_{\rm max}^{\delta}(A_1A_2E_1|BR)_\Omega$ since the two states $\ket{\psi}^{ABR}$ and $\ket{\Omega}^{A_1A_2E_1BR}$ are related by an isometry $U^{A\to A_1A_2E_1}_\cE$ (Lemma \ref{mincond}). The final equality follows from the duality relation (\ref{EQ_dual}) and the fact that $\psi^{ABR}$ is pure.

Therefore, by choosing $\eps'=\eps'''=\eps$ and $2\eps''=\eps+\sqrt{\kappa}$, we have
\begin{equation}
q^{(1)}_\eps = \log |A_2| \geq \frac{1}{2}\left[H_{\min}^{\eps}(A)_\psi- H_{\min}^{3\eps+3\sqrt{\kappa}}(A|R)_\psi \right]-\log\frac{\sqrt{2}}{\eps}.
\end{equation}
We can also obtain an upper bound for entanglement gain. We start with
\begin{eqnarray}
\fl
H_{\rm min}^{\eps'+2\eps''+\eps'''+ 2\sqrt{\kappa}}(A|R)_\psi &=& H_{\rm min}^{\eps'+2\eps''+\eps'''+ 2\sqrt{\kappa}}(A_1A_2E_1|R)_\Omega \nonumber \\
&\geq& H_{\rm min}^{\eps'''+2\sqrt{\kappa}}(A_1E_1|R)_\Omega+H_{\rm min}^{\eps''}(A_2|A_1E_1R)_\Omega -\log\frac{2}{\eps'^2} \nonumber \\
&\geq& H_{\rm min}^{\eps'''}(A_1E_1)_\Omega+H_{\rm min}^{\eps''}(A_2|A_1E_1BR)_\Omega -\log\frac{2}{\eps'^2} \nonumber \\
&=& H_{\rm min}^{\eps'''}(A_1E_1)_\Omega-H_{\rm max}^{\eps''}(A_2)_\Omega -\log\frac{2}{\eps'^2} \nonumber \\
&\geq& H_{\rm min}^{\eps'''}(A_1E_1)_\Omega-q^{(1)}_\eps-\log\frac{2}{\eps'^2}.\label{EQ_conv060}
\end{eqnarray}
The first equality holds because the systems $A$ and $A_1A_2E_1$ are related by an isometry (Lemma \ref{mincond}). The first inequality follows from the chain rule for smooth min-entropy (Lemma \ref{chain}). The second inequality follows from Lemma~\ref{LEM_Dec} and Lemma~\ref{LEM_COND}. The second equality follows from the duality relation (\ref{EQ_dual}) and the fact that $\Omega^{A_1A_2E_1RB}$ is a pure state. The last inequality holds because
\[
q^{(1)}_\eps=\log|A_2| \geq H_{\rm max}^{\eps''}(A_2)_\Omega.
\]
Let us choose $\eps'''=\sqrt{\kappa}$. We can then find a lower bound for $H_{\rm min}^{\eps'''}(A_1E_1)_\Omega$ on the right-hand side of (\ref{EQ_conv060}) as follows:
\begin{eqnarray}
H_{\rm min}^{\eps'''}(A_1E_1)_\Omega= H^{\sqrt{\kappa}}_{\rm min}(A_1E_1)_\Omega &=&\max_{\overline{\sigma}^{A_1E_1}\in\cB^{\sqrt{\kappa}}(\Omega^{A_1E_1})}H_{\rm min}(A_1E_1)_{\overline{\sigma}^{A_1E_1}}\nonumber\\
&\geq& H_{\rm min}(A_1E_1)_{\tau^{A_1}\otimes\varphi^{E_1}}\nonumber\\
&=& H_{\rm min}(A_1)_{\tau^{A_1}}+H_{\rm min}(E_1)_{\varphi^{E_1}}\nonumber\\
&\geq&\log |A_1| \nonumber\\
&=& e^{(1)}_\eps. \label{EQ_conv06}
\end{eqnarray}
The first inequality follows because \reff{EQ_conv02} implies that
$\|\tau^{A_1}\otimes\varphi^{E_1}-\Omega^{A_1E_1}\|_1\leq \kappa $,
which in turn implies that $\tau^{A_1}\otimes\varphi^{E_1}\in\cB^{\sqrt{\kappa}}(\Omega^{A_1E_1})$. The last inequality follows because $H_{\rm min}(E_1)_{\varphi^{E_1}}$ is non-negative, since $\varphi^{E_1}$ is a state. Putting (\ref{EQ_conv060}) and (\ref{EQ_conv06}) together and choosing $\eps'=\eps''=\eps$, we have
\begin{eqnarray}
e^{(1)}_\eps &\leq& q^{(1)}_\eps + H_{\rm min}^{3\eps+ 3\sqrt{\kappa}}(A|R)_\psi +\log\frac{2}{\eps^2}. 
\end{eqnarray}
\end{proof}

%%%%%%%%%%%%%%%%%%%%%%%%%
% Conclusion
%%%%%%%%%%%%%%%%%%%%%%%%%
\section{Conclusions and Discussions}
In this paper we obtain bounds on the quantum communication cost and the entanglement gain for the one-shot FQSW in terms of smooth min- and max- entropies. The one-shot FQSW can be considered to be at the apex of the existing family tree of protocols since it yields the optimal rates of the (asymptotic) FQSW, which in turn is known to be the mother of all protocols in this tree. We also employ our one-shot results to explicitly prove the optimality of the asymptotic rates. We introduce a resource inequality framework in the one-shot regime which yields achievable rates for the children protocols of the one-shot FQSW. 
%We also establish the optimality of these rates for the one-shot mother and fa%ther protocols and 
We also obtain bounds on the one-shot quantum capacity of a noisy channel in terms of the {\em{same}} entropic quantity, namely a smooth 
conditional max-entropy, unlike previously obtained bounds \cite{BD10IT}.

Note that the entropic quantities characterizing the upper and lower bounds on the quantum communication cost for the one-shot FQSW are different. The reason behind this can be understood through the following examples pointed out to us by Berta et al \cite{zurich}. For the case in which Alice and Bob initially share a state which is not correlated with the reference, perfect state transfer can be achieved without any quantum communication. This agrees with the lower bound on the quantum communication cost (in Theorem \ref{thmconverse}), which (modulo the $\eps$-dependent factor) vanishes for such a state. In contrast, there exists examples of genuine tripartite entangled states (in which Alice's system is classically correlated with the reference) for which one requires the quantum communication cost to be of the order of the dimension of the Hilbert space of Alice's system. This in turn agrees with the upper bound on the quantum communication cost in Theorem~\ref{theorem5}.

\section*{Acknowledgments}
We are very grateful to M.~M.~Wilde for carefully reading an earlier version of this manuscript and pointing out an error in its Appendix. We would like to thank K.~Audenaert, R.~Colbeck, M.~Mosonyi, and J.~Oppenheim for helpful discussions. We would also like to thank M.~Berta, M.~Christandl and R.~Renner whose insightful comments and examples helped us appreciate the complexity of the problem and correct some errors. The research leading to these results has received funding from the European Community's Seventh Framework Programme (FP7/2007-2013) under grant agreement number 213681.

\appendix
%%%%%%%%%%%%%%%%
% Lemmas
%%%%%%%%%%%%%%%%
\section{Useful Lemmas}\label{APPENDIX_A}
\begin{lemma}\label{LEM_IID} For $\eps>0$, and $\rho^A\in\cD(\cH_A)$, we have
\begin{eqnarray*}
\lim_{\eps \to 0}\lim_{n\to \infty}\frac{1}{n} H_{\rm 0}^\eps(A)_{\rho^{\otimes n}}&=& H(A)_\rho.
\end{eqnarray*}
\end{lemma}
\begin{proof}
To show that 
\begin{equation}\label{EQ_UP00}
\lim_{\eps \to 0}\lim_{n\to \infty}\frac{1}{n} H_{\rm 0}^\eps(A)_{\rho^{\otimes n}}\leq H(A)_\rho
\end{equation}
we resort to the following more general inequality, for a sequence of states $\widehat{\rho}^A:=\{\rho_n^A\}_{n=1}^\infty$
\begin{equation}\label{EQ_sup_ent01}
\lim_{\eps \to 0}\lim_{n\to \infty}\frac{1}{n} H_{\rm 0}^\eps(A)_{\rho_n}\leq \overline{S}(A)_{\widehat{\rho}}.
\end{equation}
In the above, $\overline{S}(A)_{\widehat{\rho}}$ denotes the \emph{sup-spectral entropy rate} which is defined as
\[
\overline{S}(A)_{\widehat{\rho}}:=\inf\left\{\gamma:\liminf_{n\to\infty}\tr \left[P_n^\gamma\rho_n^A\right]=1\right\}
\]
where $P_n^\gamma$ is a projector defined as
\[
P_n^\gamma:=\{\rho_n^A\geq 2^{-n\gamma}\bbI_{A_n}\}.
\]
Equation \reff{EQ_sup_ent01} holds because for every $\gamma\geq \overline{S}(A)_{\widehat{\rho}}$ and every $\delta>0$ there exists a positive integer $n_0$ such that for every $n\geq n_0$
\[
\tr\left[P_n^\gamma \rho_n^A\right]\geq 1-\delta.
\]
Let $\widetilde{\rho}_n^A:=P_n^\gamma \rho_n^A P_n^\gamma$. The Gentle Measurement lemma (Lemma~\ref{gentle}) gives
\[
\|\widetilde{\rho}_n^A-\rho_n^A\|_1\leq 2\sqrt{\delta}.
\]
Equivalently, $\widetilde{\rho}_n^A\in\cB^{\delta'}(\rho_n^A)$, where $\delta'=\sqrt{4\sqrt{\delta}-4\delta}$. Then
\begin{eqnarray}
H_0^{\delta}(A)_{\rho_n}&\leq& H_0(A)_{\widetilde{\rho}_n} \nonumber \\
&=& \log\tr\Pi_{\widetilde{\rho}_n} \nonumber \\
&\leq& \log\tr P_n^\delta \nonumber \\
&=& \log \tr [\{\rho_n^A\geq 2^{-n\gamma}\bbI_{A_n}\}\bbI_{A_n}]\nonumber \\
&\leq& n\gamma,
\end{eqnarray}
and therefore \reff{EQ_sup_ent01} holds. When we restrict our considerations to a sequence $\widehat{\rho}^A:=\{\rho^{\otimes n}\}_{n=1}^\infty$, $\overline{S}(A)_{\widehat{\rho}}=H(A)_\rho$ \cite{Bowen2}, and we obtain \reff{EQ_UP00}.

The opposite inequality 
\begin{equation}\label{EQ_UP01}
\lim_{\eps \to 0}\lim_{n\to \infty}\frac{1}{n} H_{\rm 0}^\eps(A)_{\rho^{\otimes n}}\geq H(A)_\rho
\end{equation}
follows directly from
\[
H_0^{2\sqrt{\eps}}(A)_{\rho}\geq \min_{\overline{\rho}\in\cB^{2\sqrt{\eps}}(\rho^A)}H_0(A)_{\overline{\rho}},
\]
and
\[
\lim_{\eps \to 0}\lim_{n\to \infty}\frac{1}{n}\left[\min_{\overline{\rho}\in\cB^{2\sqrt{\eps}}((\rho^A)^{\otimes n})}H_0(A)_{\overline{\rho}}\right]=H(A)_\rho.
\]
\end{proof}

We make use of the following properties of the min- and max-entropies which were proved in \cite{TCR10IEEEIT, Dupuis2010}:
\begin{lemma}
\label{mincond}
Let $0<\eps \leq 1$, $\rho^{AB} \in \cD_{\leq} (\cH_{AB})$, and let $U^{A\rightarrow C}$ and $V^{B\rightarrow D}$ be two isometries with $\omega^{CD}:= (U \otimes V) \rho^{AB}(U^\dagger \otimes V^\dagger)$, then
\begin{eqnarray*}
H_{\min}^\eps(A|B)_\rho&=&H_{\min}^\eps(C|D)_\omega\\
H_{\max}^\eps(A|B)_\rho&=&H_{\max}^\eps(C|D)_\omega.
\end{eqnarray*}
\end{lemma}

\begin{lemma}[Data-processing inequality for smooth max-entropy]\label{data}\cite{TCR10IEEEIT}
Let $0\leq \eps\leq 1$ and $\rho^{AB}\in\cD(\cH_{AB})$, and let
$\cE^{B\rightarrow C}$ be a CPTP map with $\sigma^{AC}:= ({\rm{id}}_A
\otimes \cE^{B\to C})\rho^{AB}$. Then
\[
H_{\rm max}^{\eps}(A|B)_\rho\le H^{\eps}_{\rm max}(A|C)_\sigma.
\]
\end{lemma}

\begin{lemma}[Chain rule for smooth min- and max-entropies]\label{chain}\cite{Dupuis2010}
Let $0<\eps\leq 1$, $\eps',\eps''\geq0$ and $\rho^{ABC}\in\cD(\cH_{ABC})$. Then
\[
H_{\rm min}^{\eps+2\eps'+\eps''}(AB|C)_\rho\geq H^{\eps'}_{\rm min}(A|BC)_\rho+H_{\rm min}^{\eps''}(B|C)_\rho-\log\frac{2}{\eps^2}.
\]
Let $\varphi^{ABCR}$ be a purification of $\rho^{ABC}$. Furthermore,  through the duality relation (\ref{EQ_dual}), we have
\[
H_{\rm max}^{\eps'}(A|R)_\varphi\geq H^{\eps+2\eps'+\eps''}_{\rm max}(AB|R)_\varphi - H_{\rm max}^{\eps''}(B|AR)_\varphi-\log\frac{2}{\eps^2}.
\]
\end{lemma}

\begin{lemma}\cite{TCR10IEEEIT}\label{LEM_COND}
Let $0<\eps\leq 1$ and $\rho^{ABC}\in\cD_{\leq}(\cH_{ABC})$. Then
\begin{eqnarray*}
H_{\rm min}^\eps(A|BC)_\rho &\leq& H_{\rm min}^\eps(A|B)_\rho \\
H_{\rm max}^\eps(A|BC)_\rho &\leq& H_{\rm max}^\eps(A|B)_\rho.
\end{eqnarray*}
\end{lemma}

\begin{lemma}\cite{Dupuis2010}\label{LEM_DIM}
Let $0<\eps\leq 1$ and $\rho^{ABC}\in\cD_{\leq}(\cH_{ABC})$. Then
\begin{eqnarray*}
H_{\rm min}^\eps(AB|C)_\rho &\leq& H_{\rm min}^\eps(A|C)_\rho +\log|B|.
\end{eqnarray*}
Let $\varphi^{ABCR}$ be a purification of $\rho^{ABC}$. Furthermore,  through the duality relation (\ref{EQ_dual}), we have
\[
-H_{\rm max}^\eps(A|BR)_\varphi \geq - H_{\rm max}^\eps(AB|R)_\varphi-\log|B|.
\]
\end{lemma}

\begin{lemma}{\cite{roger}}\label{berta}
For any $0<\eps\leq 1$, and $\rho^{AB}\in\cD(\cH_{A}\otimes\cH_B)$,
\be\label{EQ_max_0}
H_{\rm max}^\eps(A|B)_\rho \leq {\widetilde{H}}_0^\eps(A|B)_\rho,
\ee
where ${\widetilde{H}}_0^\eps(A|B)_\rho$ is defined through \reff{ent_Ho_smooth}.
\end{lemma}
\begin{proof}
Let $\omega^{AB}\in\cB^\eps(\rho^{AB})$ be an operator for which the minimum in \reff{ent_Ho_smooth} is achieved, and let $\omega^{ABE}$ be its purification. Then
\begin{eqnarray}\label{pf}
H_{\rm max}^\eps(A|B)_\rho &=& \min_{\overline{\rho}^{AB}\in\cB^\eps(\rho^{AB})}H_{\rm max}(A|B)_{\overline{\rho}} \nonumber \\
&\leq& H_{\rm max}(A|B)_\omega \nonumber \\
&=& - H_{\rm min}(A|E)_\omega\nonumber\\
&\le & \min_{\nu^E} \left[- H_{\rm min}(\omega^{AE}|{\nu^E})\right]\nonumber\\
&\le & - H_{\rm min}({\omega^{AE}}|{\omega^E})\nonumber\\
&=& {{H}}_0(A|B)_\omega \nonumber \\
&=& {\widetilde{H}}_0^\eps(A|B)_\rho,
\end{eqnarray}
where $H_{\rm min}(\omega^{AE}|{\nu^E})= - D_{\max}(\omega^{AE}|| \bbI_A \otimes \nu^E)$. The second equality in \reff{pf} follows from the duality relation \reff{EQ_dual} and the third equality follows from Proposition 3.1 of \cite{BertaThesis} (extended to subnormalized states), which states that for any tripartite pure state $\omega^{ABE}$, $H_{\rm min}({\omega^{AE}}|{\omega^E})=-H_0(A|B)_\omega$.
\end{proof}

\section{Proof of One-shot Decoupling Theorem}\label{Appendix_B}
Here we provide a proof of the decoupling theorem, Theorem~\ref{thm_decoupling}, for sake of completeness. Various versions of the proof can be found in, e.g., \cite{BD10IT,BCR09,DupuisThesis,Dupuis2010}.

\begin{proof}
For any fixed $0<\eps\leq 1$, let $\overline{\rho}^{AR}\in\cB^{\eps}(\rho^{AR})$ and let
\[
\overline{\sigma}^{A_1R}(U)=\tr_{A_2} \left[(U\otimes\bbI_R)\overline{\rho}^{AR}(U\otimes\bbI_R)^\dagger \right].
\]
Since the unitary operator only acts on the system $A$, we have that $ \overline{\sigma}^R(U)=\overline{\rho}^R.$
Similarly, taking the partial trace over $A_1$ on both sides of (\ref{EQ_dir_one-shot00}) yields $\sigma^R(U)=\rho^R.$
By the triangle inequality,
\begin{eqnarray}
\fl
\|\sigma^{A_1R}(U)-\tau^{A_1}\otimes\rho^R\|_1\nonumber\\
\leq \|\sigma^{A_1R}(U)-\overline{\sigma}^{A_1R}(U)\|_1+\|\overline{\sigma}^{A_1R}(U)-\tau^{A_1}\otimes\overline{\rho}^R\|_1
+\|\tau^{A_1}\otimes\overline{\rho}^R-\tau^{A_1}\otimes\rho^R\|_1 \nonumber \\
=\|\sigma^{A_1R}(U)-\overline{\sigma}^{A_1R}(U)\|_1+\|\overline{\sigma}^{A_1R}(U)-\tau^{A_1}\otimes\overline{\rho}^R\|_1
+\|\overline{\rho}^R-\rho^R\|_1 \nonumber \\
\leq\|\overline{\sigma}^{A_1R}(U)-\tau^{A_1}\otimes\overline{\rho}^R\|_1+2 \|\overline{\sigma}^{A_1R}(U)-\sigma^{A_1R}(U)\|_1, \label{EQ_dir_one-shot03}
\end{eqnarray}
where the last inequality follows from the monotonicity of the trace distance under partial trace:
\[
\|\overline{\rho}^R-\rho^R\|_1=\|\overline{\sigma}^R(U)-\sigma^R(U)\|_1\leq \|\overline{\sigma}^{A_1R}(U)-\sigma^{A_1R}(U)\|_1.
\]
Now by Lemma~3.2 of \cite{HHYW08}, we have
\begin{eqnarray}
\int_{U(A)}\|\sigma^{A_1R}(U)-\overline{\sigma}^{A_1R}(U)\|_1~dU&\leq& \|\overline{\rho}^{A_1R}-\rho^{A_1R}\|_1 \nonumber\\
&\leq& 2\eps \label{EQ_dir_one-shot04}
\end{eqnarray}
where the last inequality follows because $\overline{\rho}^{A_1R}\in\cB^\eps(\rho^{A_1R})$.
Substituting (\ref{EQ_dir_one-shot04}) into (\ref{EQ_dir_one-shot03}) yields
\begin{equation}\label{EQ_dir_one-shot05}
\int_{U(A)}\|\sigma^{A_1R}(U)-\tau^{A_1}\otimes\rho^R\|_1~dU\leq \int_{U(A)}\|\overline{\sigma}^{A_1R}(U)-\tau^{A_1}\otimes\overline{\rho}^R\|_1~dU+4\eps.
\end{equation}
Next we use the following inequalities (Lemma~5.1.3 of \cite{RennerThesis}): Let $H$ be a Hermitian operator in $\cB(\cH)$, and $\Omega\in\cB(\cH)$. Then
\begin{eqnarray}
\|H\|_1 &\leq& \sqrt{\tr\Omega} \|\Omega^{-1/4}H\Omega^{-1/4}\|_2\\
\|H\|_1^2 &\leq& \tr\Omega \|\Omega^{-1/4}H\Omega^{-1/4}\|_2^2. \label{EQ_dir_one-shot06}
\end{eqnarray}
Hence
\begin{eqnarray}
\fl
\int_{U(A)}\|\overline{\sigma}^{A_1R}(U)-\tau^{A_1}\otimes\overline{\rho}^R\|_1~dU \nonumber\\
\leq \int_{U(A)} \sqrt{\tr\Omega}\sqrt{\|\Omega^{-1/4}(\overline{\sigma}^{A_1R}(U)
-\tau^{A_1}\otimes\overline{\rho}^R)\Omega^{-1/4}\|_2^2}~dU \nonumber \\
\leq \sqrt{\tr\Omega}\sqrt{\int_{U(A)}\|\Omega^{-1/4}(\overline{\sigma}^{A_1R}(U)-\tau^{A_1}\otimes\overline{\rho}^R)\Omega^{-1/4}\|_2^2~dU}.
\label{EQ_dir_one-shot07}
\end{eqnarray}
The first inequality follows from (\ref{EQ_dir_one-shot06}). The second inequality follows from the concavity of the function $f(x)=\sqrt{x}$.
Choose $\Omega=\bbI_{A_1}\otimes\omega_R$, where $\omega_R\in\cD(\cH_R)$, and let
\begin{eqnarray}
\widetilde{\sigma}^{A_1R}(U)&=\Omega^{-1/4}\overline{\sigma}^{A_1R}(U)\Omega^{-1/4}\nonumber \\
&= \left(\bbI_{A_1}\otimes\omega_R^{-1/4}\right)\overline{\sigma}^{A_1R}(U)\left( \bbI_{A_1}\otimes\omega_R^{-1/4}\right)\nonumber \\
&= \left(\bbI_{A_1}\otimes\omega_R^{-1/4}\right)\tr_{A_2} \left[(U\otimes\bbI_R)\overline{\rho}^{AR}(U\otimes\bbI_R)^\dagger \right]
\left( \bbI_{A_1}\otimes\omega_R^{-1/4}\right)\nonumber \\
&= \tr_{A_2} \left[(U\otimes\bbI_R)\left(\bbI_{A_1}\otimes\omega_R^{-1/4}\right)\overline{\rho}^{AR}
\left( \bbI_{A_1}\otimes\omega_R^{-1/4}\right)(U\otimes\bbI_R)^\dagger \right]\nonumber \\
&= \tr_{A_2} \left[(U\otimes\bbI_R)\widetilde{\rho}^{AR}(U\otimes\bbI_R)^\dagger \right],\label{EQ_dir_one-shot08}
\end{eqnarray}
with $\widetilde{\rho}^{AR}:=\left(\bbI_{A_1}\otimes\omega_R^{-1/4}\right)\overline{\rho}^{AR}\left( \bbI_{A_1}\otimes\omega_R^{-1/4}\right)$ in the last equality. Note that $\widetilde{\rho}^{R}=\tr_A\widetilde{\rho}^{AR}= \omega_R^{-1/4}\overline{\rho}^{R}\omega_R^{-1/4}$.
Continuing from (\ref{EQ_dir_one-shot07}),
\begin{eqnarray}
\fl
\int_{U(A)}\|\Omega^{-1/4}(\overline{\sigma}^{A_1R}(U)-\tau^{A_1}\otimes\overline{\rho}^R)\Omega^{-1/4}\|_2^2~dU
\nonumber\\
=\int_{U(A)}\|\widetilde{\sigma}^{A_1R}(U)-\tau^{A_1}\otimes\widetilde{\rho}^R\|_2^2~dU\nonumber\\
=\int_{U(A)} \tr \left[\left(\widetilde{\sigma}^{A_1R}(U)-\tau^{A_1}\otimes\widetilde{\rho}^R\right)^2 \right]~dU\nonumber\\
= \int_{U(A)} \left(\tr\left[\widetilde{\sigma}^{A_1R}(U)^2\right]- 2\tr\left[\widetilde{\sigma}^{A_1R}(U)(\tau^{A_1}\otimes\widetilde{\rho}^R)\right]
+\tr\left[(\tau^{A_1}\otimes\widetilde{\rho}^R)^2\right] \right)~dU\nonumber\\
= \int_{U(A)}\tr\left[\widetilde{\sigma}^{A_1R}(U)^2\right]~dU - \tr\left[(\tau^{A_1}\otimes\widetilde{\rho}^R)^2\right] \nonumber\\
= \int_{U(A)}\tr\left[\widetilde{\sigma}^{A_1R}(U)^2\right]~dU - \frac{1}{|A_1|}\tr\left[(\widetilde{\rho}^R)^2\right] \nonumber\\
\leq \frac{1}{|A_2|}\tr\left[(\widetilde{\rho}^{AR})^2\right]. \label{EQ_dir_one-shot09}
\end{eqnarray}
The last inequality follows from Lemma~C.2 of \cite{BCR09} which states that
\[
\int_{U(A)}\tr\left[\widetilde{\sigma}^{A_1R}(U)^2\right]\ dU\leq \frac{1}{|A_1|}\tr\left[(\widetilde{\rho}^R)^2\right]
+\frac{1}{|A_2|}\tr\left[(\widetilde{\rho}^{AR})^2\right].
\]
Let $\overline{\rho}^{AR}\in\cD_{\leq}(\cH_{AR})$ and $\omega_R\in\cD(\cH_R)$ be such that
\begin{eqnarray}
H_{\rm min}^\eps(A|R)_\rho &=& - D_{\rm max}(\overline{\rho}_{AR}||\bbI_A\otimes\omega_R) \nonumber\\
&=& H_{\rm min}(A|R)_{\overline{\rho}|\omega} \nonumber\\
&\leq& H_{\rm C}(A|R)_{\overline{\rho}|\omega}, \label{CL}
\end{eqnarray}
where $H_{\rm C}(A|R)_{\overline{\rho}|\omega}$ is the quantum collision entropy of $\overline{\rho}_{AR}$ relative to $\omega_R$:
\begin{eqnarray}
H_{\rm C}(A|R)_{\overline{\rho}|\omega}&:=& - \log \tr\left[\left((\bbI_A\otimes\omega_R^{-1/4}) \overline{\rho}^{AR}(\bbI_A\otimes\omega_R^{-1/4}) \right)^2  \right] \nonumber\\
&=&-\log\tr\left[(\widetilde{\rho}^{AR})^2\right].
\end{eqnarray}
The last inequality in \reff{CL} follows from Lemma~B.16 of \cite{BCR09}. 
Now suppose we choose
\begin{eqnarray}
\log |A_1| &\leq& \frac{1}{2}\bigl[\log|A|+H_{\rm min}^\eps(A|R)_\rho\bigr]+ \log\eps\nonumber \\
&\leq&  \frac{1}{2}\bigl[\log|A|+H_{\rm C}(A|R)_{\overline{\rho}|\omega}\bigr]+ \log\eps\nonumber \\
&=&\frac{1}{2}\Bigl[\log|A|-\log\tr\bigl[(\widetilde{\rho}^{AR})^2\bigr]\Bigr]+\log{\eps}. \label{EQ_dir_one-shot10}
\end{eqnarray}
Equation (\ref{EQ_dir_one-shot10}) then implies that
\begin{eqnarray}\label{EQ_dir_one-shot11}
\log\left(\frac{|A_1|}{\eps}\right)^2 &\leq& \log|A| -\log\tr\left[(\widetilde{\rho}^{AR})^2\right]\nonumber\\
&=& \log \frac{|A|}{\tr\left[(\widetilde{\rho}^{AR})^2\right]}.
%\Rightarrow \frac{\tr\left[(\widetilde{\rho}^{AR})^2\right]}{|A|}&\leq& \frac{\eps^2}{|A_1|^2}.
\end{eqnarray}
From (\ref{EQ_dir_one-shot05}), (\ref{EQ_dir_one-shot07}) and (\ref{EQ_dir_one-shot09}) we obtain
\begin{eqnarray}
\int_{U(A)}\|\sigma_{A_1R}(U)-\tau_{A_1}\otimes\rho_R\|_1\ dU &\leq& 4\eps+\sqrt{|A_1|}\sqrt{ \frac{1}{|A_2|}\tr\left[(\widetilde{\rho}_{AR})^2\right]}\nonumber\\
&=&4\eps+|A_1| \sqrt{ \frac{1}{|A|}\tr\left[(\widetilde{\rho}_{AR})^2\right]}\nonumber \\
%&\leq& 4\eps+|A_1| \sqrt{ \frac{1}{H_{\rm max}^\eps(A)_\rho}\tr\left[(\widetilde{\rho}_{AR})^2\right]}\nonumber \\
&\leq& 4\eps+ |A_1|\sqrt{\frac{\eps^2}{|A_1|^2}}\nonumber\\
&\leq& 5\eps,
\end{eqnarray}
where the last inequality follows from (\ref{EQ_dir_one-shot11}).
\end{proof}

\end{document}